\documentclass[aps,prb,twocolumn,amsmath,amssymb]{revtex4-2}
\usepackage{mathtools}

\usepackage{braket}
\usepackage{graphicx, epsfig}
\usepackage{overpic}
\usepackage{bm}
\usepackage{dcolumn}
\usepackage{xcolor}
\usepackage{mathtools}
\usepackage{float}
\usepackage{tikz}
\usepackage{csquotes}
\usepackage[normalem]{ulem}
\usepackage[caption=false]{subfig}
\usetikzlibrary{arrows.meta}

\usepackage{amsmath,amssymb,amsbsy,amstext,mathrsfs}
\usepackage{MnSymbol}
\DeclareSymbolFontAlphabet{\mathscrsfs}{rsfs}
\usepackage[breaklinks=true,colorlinks,citecolor=blue,linkcolor=blue,urlcolor=blue]{hyperref}

\usepackage[capitalize]{cleveref}

\def\noi{\noindent}
\def\bc{\begin{center}}
\def\ec{\end{center}}
\newcommand{\bea}{\begin{equation}}
\newcommand{\eea}{\end{equation}\noi}
\newcommand{\ber}{\begin{eqnarray}}
\newcommand{\eer}{\end{eqnarray}\noi}

\begin{document}	
	
\title{Fermions on a 1D lattice: localized sources and sinks with dephasing}
	
\author{Elka Bhattacharya}\email{belka@tifrh.res.in}
\author{Sushant Saryal}
%\author{Elka Bhattacharya}\email{belka@tifrh.res.in}
\author{Soumya Ghosh}%\email{soumya.ghosh@tifrh.res.in}
\author{Kabir Ramola}\email{kramola@tifrh.res.in}
\affiliation{Tata Institute of Fundamental Research, Hyderabad 500046, India}

\date{\today}
	
\begin{abstract}
We study a general one-dimensional spinless and non-interacting fermionic system subject to a localized source, sink, and bulk dephasing. Within the Lindblad framework, we compute the time evolution of the density profile and spatial correlation functions. We find that the presence of bulk dephasing suppresses certain coherent quantum features, such as the Friedel oscillations, and it alters the transport dynamics to exhibit two distinct dynamical regimes instead of three as observed in the absence of dephasing. This effect can be understood as the destruction of ballistic motion caused by the dephasing noise. Under strong dephasing, the density profile becomes similar to the one expected for a classical diffusive regime. We also investigate the system in the presence of both a localized source and sink, placed at a distance of $\Delta$. Interestingly, quantum coherence generates secondary density peaks at integer multiples of the source-sink separation $\Delta$, which are systematically washed out as bulk dephasing drives the system toward the classical diffusive limit.
\end{abstract}

\maketitle
\section{Introduction}

Understanding the dynamics and thermalization of open quantum systems is a central challenge in non-equilibrium statistical mechanics and condensed matter physics~\cite{Lindblad1976, Gorini1976, Breuer2007, Rotter_2015, RevMod_Landi2022, Reichental2017}. A key class of problems concerns the transport of energy and particles in extended one-dimensional lattices that are locally coupled to baths~\cite{Levitov1993, Levitov1996, Schonhammer2007, Prosen2011,katha2024,katha2026}. Recent studies have explored the dynamics of bosonic and fermionic systems driven by localized particle injection~\cite{Butz2010, Krapivsky2019, Krapivsky2020, Trivedi2023, Ray2026} and localized particle loss~\cite{fluctuation,Ultracold,Dutta2020,Wolff2020,Diehl2021,Kollath2022,Alba2022,Kollath2023,Marco2025}. In coherent (non-dissipative) setups of non-interacting systems, the density profile spreads ballistically, and the total particle number grows linearly before saturating through equilibration with the bath. Analytical and numerical approaches, including the Redfield equation, Lindblad master equation, and quantum Langevin equation, show that this ballistic propagation obeys a universal scaling form determined solely by the underlying quantum statistics, whether fermionic or bosonic~\cite{Trivedi2023, Alba2022}.

While the unitary evolution of non-interacting systems gives rise to purely ballistic transport accompanied by coherent phenomena such as Friedel oscillations, realistic quantum systems are inevitably coupled to fluctuating environments that induce decoherence. Local dephasing is a key open-system mechanism that models the loss of phase coherence without particle or energy exchange~\cite{Esposito2005, Gaspard2005, Znidaric1_2010, Znidaric2_2010, Eisler_2011, Mendoza-Arenas_2013}. It effectively captures thermal noise, random scattering, interactions with a fluctuating background, and the back-action of continuous local measurements. By continuously monitoring local site occupations, dephasing strongly suppresses off-diagonal density-matrix elements. Consequently, dephasing fundamentally alters the transport mechanisms of the lattice, driving a dynamical crossover from quantum ballistic spreading to classical diffusive transport~\cite{Znidaric1_2010, Znidaric2_2010, Eisler_2011, Mendoza-Arenas_2013, Marco2021, Ren2024, Ray2026}. Analytically, this crossover was first demonstrated by recognizing that the many-body master equation for such non-interacting models yields a closed, decoupled hierarchy of two-point correlation functions~\cite{Znidaric2_2010}. 
%This exact framework has since been successfully extended to study steady-state full counting statistics~\cite{Znidaric_FCS_2014} and the exact time evolution~\cite{Znidaric_TimeEvol_2025} of these open quantum setups. 
Investigating this interplay between boundary-driven coherent injection or loss and bulk incoherent dephasing is crucial for understanding transport in mesoscopic setups and solid-state devices.

In this context, we analyze a one-dimensional spinless and non-interacting fermionic system with a localized sink, a localized source, and uniform dephasing. Using the Lindblad master equation, we compute the time evolution of the density profile and spatial correlations, focusing on the dephasing-driven crossover from ballistic to diffusive transport. Dephasing strongly suppresses nonlocal quantum coherence, including Friedel oscillations around the impurity. It also qualitatively alters particle loss/growth: instead of the three dynamical regimes present in the coherent case, set by ballistic propagation and boundary reflections, only two remain with dephasing, as the system-size-dependent steady current is destroyed. We analyze the steady-state configuration of a lattice coupled to a source and a sink separated by a distance $\Delta$. In the fully coherent regime, the steady-state density profile exhibits secondary interference peaks at integer multiples of $\Delta$. These peaks are a clear hallmark of quantum coherence and are progressively erased as increasing dephasing strength drives the system toward the classical limit.

The paper is organized as follows. Sec.~\ref{Model} introduces the open quantum system model, defining the tight-binding Hamiltonian and Lindblad master equation. Sec.~\ref{sink} considers a localized sink, demonstrating the suppression of Friedel oscillations, the reduction from three to two dynamical regimes, and an analytical scaling solution for spatial density in the strong-dephasing (Zeno) limit. Sec.~\ref{source} describes particle dynamics under a localized source, detailing exact correlation-matrix evolution, ballistic fermion spreading, and dephasing-induced coherence suppression. Sec.~\ref{sink_source} studies the combined localized source–sink setup, revealing coherent steady-state secondary peaks spaced by the source–sink distance $\Delta$ and their disappearance under dephasing. Finally, Sec.~\ref{conclusions} summarizes our main results and outlines future directions.

\par\vspace{0.5em}

 \section{Model}\label{Model}

%%%%%%%%%%%%%%%%%%%%%%%%%%%%%%%
\begin{figure}[t!]
        \includegraphics[width=1.0\linewidth]{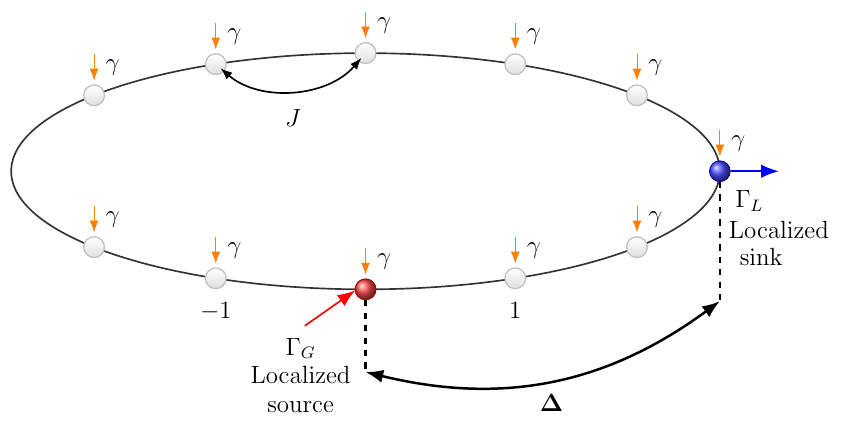}
    \caption{Schematic illustration of the source-sink with dephasing setup. A one-dimensional periodic lattice of size $L$ with nearest-neighbor hopping amplitude $J$ is coupled to a localized source at the origin and a localized sink located a distance $\Delta$ away. The source injects particles at rate $\Gamma_G$, while the sink removes particles at rate $\Gamma_L$. In addition, each lattice site is subjected to local dephasing with rate $\gamma$.} 
    \label{fig:sink_density_wod}
\end{figure}
%%%%%%%%%%%%%%%%%%%%%%%%%%%%%%%

We consider an open quantum system defined on a one-dimensional lattice with periodic boundary conditions, driven by a spatially localized fermionic source and sink in the presence of local dephasing. The lattice model is defined by the following tight-binding Hamiltonian:
\begin{align}
H = -J\sum_i\left(a^{\dagger}_ia_{i+1} + a^{\dagger}_{i+1}a_i\right)
\label{PBC_H}
\end{align}
where the lattice sites are indexed by $i \in [i_{\min}, i_{\max}]$, with $i_{\min}=-(L-1)/2$ and $i_{\max}=(L-1)/2$~\cite{Ultracold}. We assume periodic boundary conditions such that $a_{i_{\max}+1} = a_{i_{\min}}$ and $a_{i_{\min}-1} = a_{i_{\max}}$.

The time evolution of the density matrix $\rho$ of the system is governed by the Lindblad master equation~\cite{Lindblad1976,Gorini1976,Breuer2007},
\begin{align}
\partial_t \rho = -i[H,\rho]
+ \mathcal{L}_{\mathrm{sink}} \rho
+ \mathcal{L}_{\mathrm{source}} \rho
+ \mathcal{L}_{\mathrm{deph}} \rho 
\label{lindblad}
\end{align}
where $\mathcal{L}_{\mathrm{sink}} \rho$ and $\mathcal{L}_{\mathrm{source}} \rho$ are the Lindblad superoperators corresponding to the localized particle loss and localized particle gain processes, respectively, while $\mathcal{L}_{\mathrm{deph}}$ accounts for local dephasing acting uniformly on all lattice sites.

\section{Localized sink}\label{sink}
First, we consider the case of a localized sink in the absence of a source. This allows us to understand the effect of loss on the system's dynamics before introducing the source term.

\subsection{Localized sink without dephasing}
We consider the dynamics of a one-dimensional lattice model of fermions prepared at $T=0$, characterized by the Fermi momentum $k_F = \pi n_0$, where $n_0$ is the initial density~\cite{Ultracold}. Initially, fermions are distributed uniformly across all sites with a fixed density $n_0$. At time $t=0$, a local particle-loss channel (sink) is switched on at the origin ($i=0$), and the fermions are ejected from the origin at a constant rate $\Gamma_L$.

The evolution of the system is described by the following Lindblad equation, keeping ($\hbar =1)$:
\begin{align}
\partial_t \rho = -i[H,\rho]
+ \mathcal{L}_{\mathrm{sink}} \rho
\label{sink_wod}
\end{align}
where 
\begin{align}
\mathcal{L}_{\mathrm{sink}} \rho = \Gamma_L(2 a_0 \rho a^{\dagger}_0 - \{a^{\dagger}_0 a_0,\rho\}) \label{sink_diss}   
\end{align}
Eq.~\eqref{sink_wod} consists of two terms: the first term describes the unitary evolution of the system, governed by the Hamiltonian $H$ of the system [Eq.~\eqref{PBC_H}], and the second term describes the ejection of fermions induced by the local loss channel at the origin ($i=0$).

The relevant observables for the dynamics are the local density $n_i(t)$ and the total particle number $N(t) = \sum_i n_i(t)$. These observables can be analyzed using the two-point correlation functions
\begin{align}
    C_{i,j}(t) = \langle a^{\dagger}_i(t) a_j(t) \rangle = \operatorname{Tr}\Big(\rho(t) a^{\dagger}_ia_j\Big)  \label{def_corr}
\end{align}
Using the Lindblad equation \eqref{sink_wod} with the Hamiltonian \eqref{PBC_H}, the dissipator \eqref{sink_diss}, and the anticommutation relations of fermions, the equations of motion for the two-point correlation functions form a closed hierarchy that decouples from higher-order correlations~\cite{Znidaric1_2010,Znidaric2_2010}. This exact linear system can be written as~\cite{Krapivsky2019, Krapivsky2020}
\begin{align}
 \frac{d}{dt}C_{i,j} &= -iJ\big(C_{i-1,j} + C_{i+1,j} - C_{i,j+1} - C_{i,j-1}\big)  \nonumber \\
&\quad - \Gamma_L(\delta_{i,0} + \delta_{j,0})\,C_{i,j}  \label{non_homo_source_wod}
\end{align}
Since the master equation is a linear differential equation, the equation of motion for the correlation functions is also linear; hence, we can solve the coherent and incoherent evolutions separately.

Closely following Ref.~\cite{heat_transport}, we define a column vector containing all the annihilation operators of the system as
\begin{align}
  A = \{a_{-(L-1)/2}, \dots, a_{-1}, a_0, a_1, \dots, a_{(L-1)/2}\}  
\end{align}
Hence, from Eq.~\eqref{def_corr}, we can write 
\begin{align}
C = A^{\dagger} A
\end{align}
where $A^{\dagger} = \{a^{\dagger}_i\}$ is the row vector consisting of the creation operators of the system.

To obtain the coherent evolution of the correlations, we can write the Heisenberg equation of motion for the annihilation operator as
\begin{align}
    \dot{a}_j = i[H,a_j] = iJ(a_{j+1}+ a_{j-1}) \label{Hei_anni}
\end{align}
From Eq.~\eqref{Hei_anni}, we obtain a system of coupled differential equations whose coefficients are contained in the matrix $W$. Thus, for coherent evolution, we have
\begin{align}
    \dot{A}^{\dagger} = iWA^{\dagger}
\end{align}
The incoherent part can be obtained using the form of $\mathcal{L}_{\mathrm{sink}}$ alongside the anticommutation relations of fermions. Using Eq.~\eqref{sink_wod}, we get
\begin{align}
  \frac{d}{dt}\langle a^{\dagger}_i a_j \rangle = -\Gamma_L(\delta_{i,0} + \delta_{j,0}) \langle a^{\dagger}_i a_j \rangle 
\end{align}
Thus, the full evolution equation of the correlation matrix $C$ takes the form
\begin{align}
    \frac{dC}{dt} = i[W,C] + \{L,C\} \label{mat_corr_sink_wod}
\end{align}
where
\begin{align}
L_{i,j} &= -\Gamma_L \delta_{i,0} \delta_{i,j} 
\label{L_form}
\end{align}
The formal solution of Eq.~\eqref{mat_corr_sink_wod} can be written as
\begin{align}
    C(t) = e^{G \, t} C(0) e^{G^{\dagger} \, t} 
    \label{sink_sol}
\end{align}
where $G$ is a non-Hermitian matrix of the form
\begin{align}
  G = iW + L
\end{align}

\subsubsection{Local density profile}
The local density at site $i$ is given by the diagonal elements of the correlation matrix,
\begin{align}
    n_i(t) = C_{ii}(t)
    \label{diag_sink_corr_mat_wod}
\end{align}
To obtain an explicit expression for $n_i(t)$, we diagonalize $G$ and $G^\dagger$ independently as $G=V\lambda^1V^{-1}$ and $G^{\dagger}=U\lambda^2U^{-1}$, where $\lambda^1$ and $\lambda^2$ are the corresponding diagonal eigenvalue matrices. The local density can then be expressed as
\begin{align}
    n_i(t)
    =
    \sum_{\alpha,\beta}
    e^{(\lambda_{\alpha}^{1}+\lambda_{\beta}^{2})t}
    \,V_{i\alpha}\,
    X_{\alpha\beta}\,
    U^{-1}_{\beta i}
    \label{sink_density_wod}
\end{align}
where $X=V^{-1}C(0)U$ encodes the initial correlations.

The initial correlation matrix is given by
\begin{align}
C_{m,n}(0)
=
\frac{1}{L}
\sum_{k,k'}
e^{-ikm}e^{ik'n}
\left\langle
\tilde{a}^{\dagger}_k(0)\tilde{a}_{k'}(0)
\right\rangle
\label{ini_corr}
\end{align}
where the momentum-space correlation function for an initial thermal state is
\begin{align}
\left\langle
\tilde{a}^{\dagger}_k(0)\tilde{a}_{k'}(0)
\right\rangle
=
\delta_{k,k'}
\frac{1}{e^{\beta(\varepsilon_k-\mu)}+1}.
\label{initial_corr}
\end{align}
Here, $\beta = 1/(k_B T)$ is the inverse temperature, $\mu$ is the chemical potential, and $\varepsilon_k=-2J\cos k$ is the single-particle dispersion relation of the lattice.

The density profiles at different times for two values of the sink rate are presented in Figs.~\ref{fig:sink_density_wod}(a) and \ref{fig:sink_density_wod}(c). Figs.~\ref{fig:sink_density_wod}(b) and \ref{fig:sink_density_wod}(d) confirm the ballistic propagation of the density perturbation. The existence of Friedel oscillations around the loss site is clearly visible, consistent with reports in Refs.~\cite{fluctuation, Ultracold}.  Friedel oscillations can be characterized by their spatial frequency, which is directly proportional to the density,
\begin{align}
    \omega_{\mathrm{FO}} = 2 k_F = 2 \pi n_0
\end{align}
as shown in the Supplemental Material.

%%%%%%%%%%%%%%%%%%%%%%%%%%%%%%%%%%%%%%%%%%%%%
\begin{figure}[!t]
\centering
    
    \includegraphics[width=0.49\columnwidth]{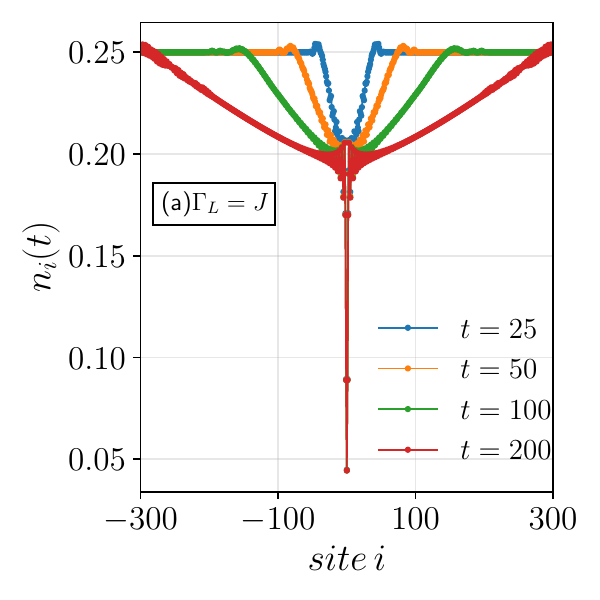}
    \includegraphics[width=0.49\columnwidth]{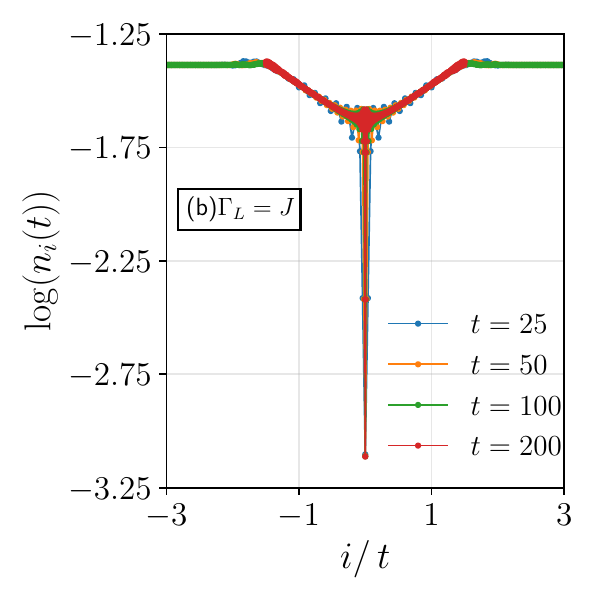}

    \includegraphics[width=0.49\columnwidth]{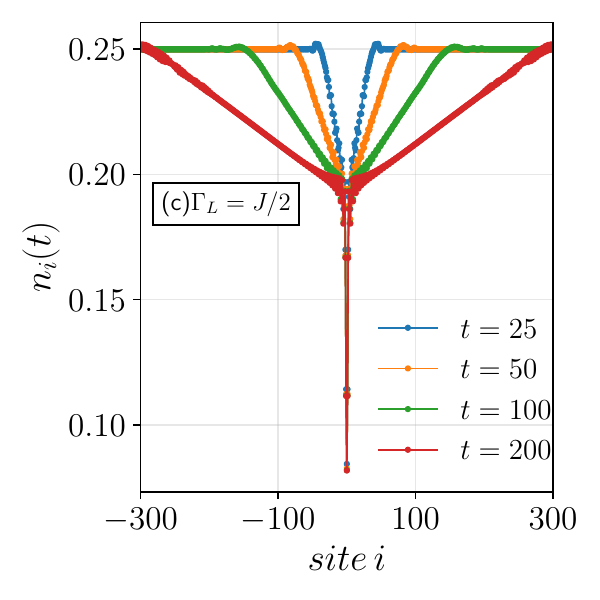}
    \includegraphics[width=0.49\columnwidth]{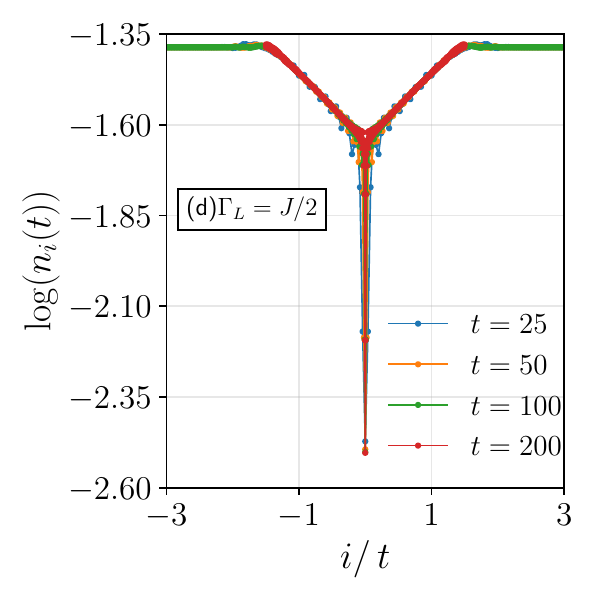}

    \caption{Local particle density profile $n_i(t)$ obtained from Eq.~\eqref{sink_density_wod} for a localized sink at the origin. Panels (a) and (c) show the results for $\Gamma_L=J$ and $\Gamma_L=J/2$, respectively, with $J=1$ and $L=601$. The scaling plots in (b) and (d) confirm the ballistic propagation of the density perturbation.}
    \label{fig:sink_density_wod}
\end{figure}
%%%%%%%%%%%%%%%%%%%%%%%%%%%%%%%%%%%%%%%%%%%%%

\subsubsection{Different time regimes}
We now consider the total particle loss rate, $|dN(t)/dt|$, where the total number of particles is given by
\begin{align}
    N(t)=\sum_i C_{ii}(t).
\end{align}
Taking the time derivative and using Eq.~\eqref{sink_sol}, we obtain
\begin{align}
    \frac{dN}{dt}
    =
    \sum_i \dot{C}_{ii}(t)
    =
    \sum_i
    \left[
    GC + CG^{\dagger}
    \right]_{ii}(t).
    \label{loss_rate}
\end{align}

\begin{figure}[t!]
\centering
   \includegraphics[width=0.85\linewidth]{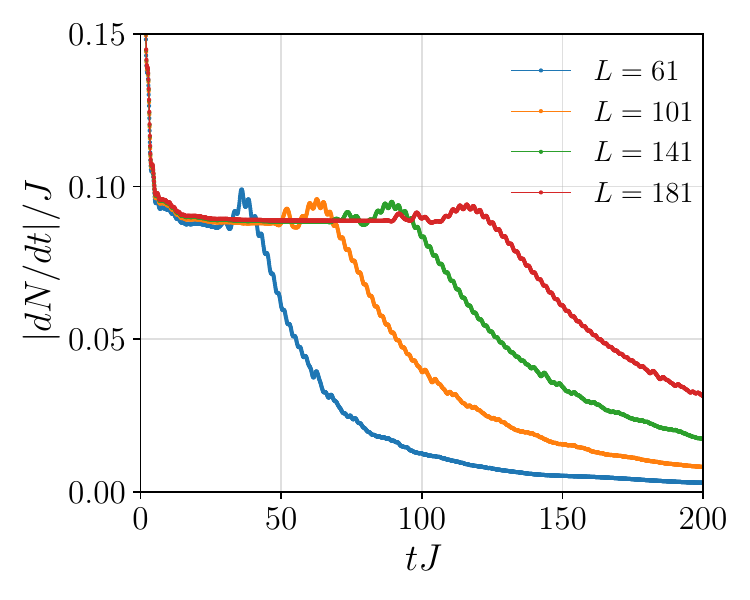}  
    \caption{Total particle loss rate, $|dN/dt|/J$ [Eq.~\eqref{loss_rate}], as a function of time $tJ$ for different system sizes $L$, with a localized sink at the origin ($i=0$) and loss rate $\Gamma_L$. The duration of the second dynamical regime scales linearly with the system size, indicating ballistic propagation of the density perturbation. For all curves, the initial filling is $N(0)/L=0.25$, and $\Gamma_L/J=1$.}
   \label{fig:sink_rate_wod}
\end{figure}

Fig.~\ref{fig:sink_rate_wod} shows the total particle loss rate, $|dN/dt|/J$, as a function of the dimensionless time $tJ$. Here, both the particle loss rate and time are scaled by the hopping amplitude $J$, the intrinsic energy scale of the system, to yield dimensionless quantities.

%%%%%%%%%%%%%%%%%%%%%%%%%%%%%%%%%%%%%%%%%%%%%
%%%%%%%%%%%%%%%%%%%%%%%%%%%%%%%%%%%%%%%%%%%%%

The loss-rate curves exhibit three distinct dynamical regimes:
(i) At very short times, the loss rate decreases rapidly due to local particle depletion in the vicinity of the sink.
(ii) This is followed by a regime in which the loss rate remains nearly constant, during which the perturbation generated by the sink propagates ballistically through the system. Consequently, the duration of this regime scales linearly with system size, as evident from the figure.
(iii) Finally, the perturbation reaches the boundaries of the system, where particle loss becomes sensitive to finite-size effects. As the dynamics crosses over from the second to the third regime, the loss rate exhibits a transient enhancement arising from constructive interference between the clockwise- and counterclockwise-propagating ballistic wavefronts at the antipodal point, $i=\pm L/2$.

\subsection{Localized sink with dephasing}
In the presence of local dephasing, the evolution of the density matrix is governed by the Lindblad master equation
\begin{align}
    \partial_t \rho = -i[H,\rho] + \mathcal{L}_{\mathrm{sink}} \rho + \mathcal{L}_{\mathrm{deph}}\rho
    \label{density_dephase}
\end{align}
where the dephasing Lindblad superoperator acts independently on each lattice site and is given by
\begin{align}
\mathcal{L}_{\mathrm{deph}} \rho
=
\gamma \sum_i
\left(
2 \hat{n}_i \rho \hat{n}_i
-
\{\hat{n}_i^{\,2},\rho\}
\right)
\label{ope_dephase}
\end{align}
with $\gamma$ denoting the dephasing (or measurement) rate and $\hat{n}_i$ the local number operator measuring the occupation of site $i$. Again, all observables of interest can be extracted from the two-point correlation function defined in Eq.~\eqref{def_corr}.

\subsubsection{Local density profile}
Using the Lindblad equation~\eqref{density_dephase} with the Hamiltonian~\eqref{PBC_H}, the dissipator~\eqref{sink_diss}, and the dephasing operator~\eqref{ope_dephase}, the equation of motion for the two-point correlation function is given by
\begin{align}
    \frac{d}{dt}C_{i,j} &= -iJ\big(C_{i-1,j} + C_{i+1,j} - C_{i,j+1} - C_{i,j-1}\big)  \nonumber \\
&\quad - \Gamma_L(\delta_{i,0} + \delta_{j,0})\,C_{i,j} + 2\gamma(\delta_{i,j} - 1)C_{i,j}. \label{sink_wd}
\end{align}
Notice that Eq.~\eqref{sink_wd} is a linear homogeneous differential equation. Vectorizing the correlation matrix as
\begin{align}
    C_r \equiv C_{i,j}(t), \qquad r=(i-1)L+j,
    \label{mat_vec}
\end{align}
with $r\in\{1,2,\dots,L^2\}$, the equation of motion can be expressed compactly as
\begin{align}
    \dot{C}=MC,
    \label{EOM_corr_wd}
\end{align}
where $M$ is the $L^2\times L^2$ coefficient matrix corresponding to Eq.~\eqref{sink_wd}. The formal solution is therefore
\begin{align}
    C(t) = U e^{\Lambda_M t} U^{-1} C(0)
    \label{sink_sol_wd}
\end{align}
where $M$ has been diagonalized as
\begin{align}
    M = U \Lambda_M U^{-1}
\end{align}
with $\Lambda_M$ being the diagonal matrix of eigenvalues of $M$.

The local density at site $i$ is then obtained as
\begin{align}
    n_i(t)=\left[U e^{\Lambda_M t}U^{-1}C(0)\right]_{i+(i-1)L}
    \label{density_sink_wd}
\end{align}
where the index $i+(i-1)L$ corresponds to the diagonal matrix element $C_{i,i}(t)$ in the vectorized representation.

\begin{figure}[t!]
\centering
    \includegraphics[width=0.85\linewidth]{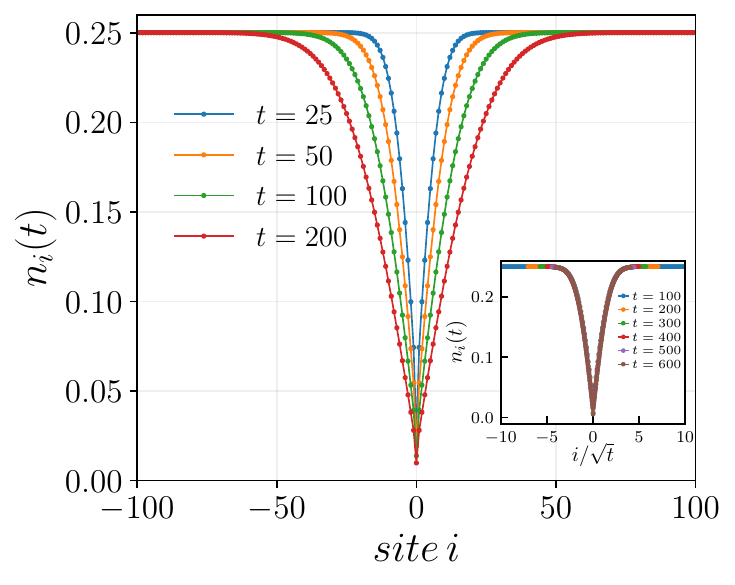}
     \caption{Local particle density profiles $n_i(t)$ [Eq.~\eqref{density_sink_wd}] at different times for a system of size $L=201$, with a localized sink at the origin ($i=0$), loss rate $\Gamma_L$, and dephasing rate $\gamma$. The inset shows the same data plotted against the scaled coordinate $i/\sqrt{t}$, demonstrating the diffusive spreading of the density perturbation. The parameters are $J=\Gamma_L=\gamma=1$, with an initial filling of $N(0)/L=0.25$.}
    \label{fig:density_sink_wd}
\end{figure}

The resulting density profiles in the presence of dephasing are shown in Fig.~\ref{fig:density_sink_wd} at different times. The dephasing bath completely suppresses the Friedel oscillations, giving rise to a smooth density profile. The inset further demonstrates that the density evolves diffusively: when plotted as a function of $i/\sqrt{t}$, the density profiles at different times collapse onto a single universal curve, confirming the expected diffusive scaling.
%%%%%%%%%%%%%%%%%%%%%%%%%%%%%%
%%%%%%%%%%%%%%%%%%%%%%%%%%%%%%

\subsubsection{Different time regimes}
\begin{figure}[t!]
\centering
\includegraphics[width=0.85\linewidth]{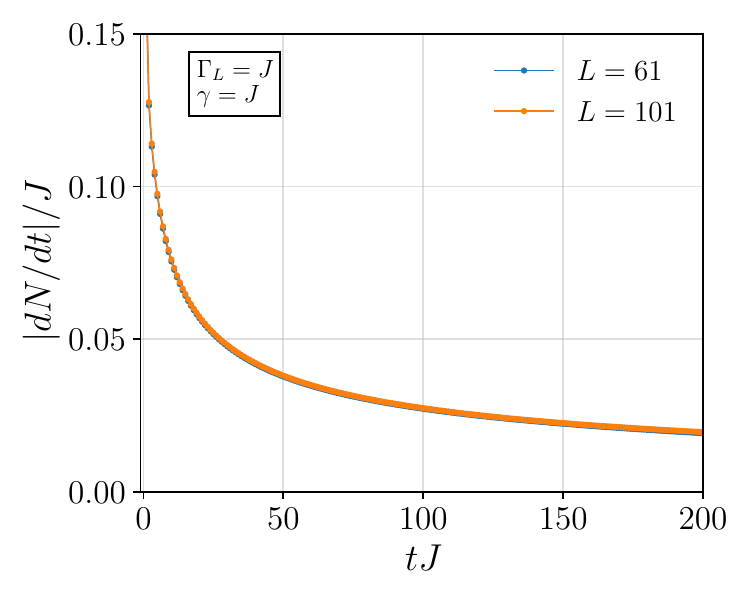}
\caption{Total particle loss rate, $|dN/dt|/J$ [Eq.~\eqref{loss_rate_wd}], as a function of time for different system sizes $L$. In the presence of dephasing, ballistic transport is destroyed, reducing the three dynamical regimes observed in the coherent case to two. Consequently, the curves for different system sizes collapse onto a single curve. The parameters are $J=\Gamma_L=\gamma=1$, and the initial filling is $N(0)/L=0.25$.} 
\label{fig:sink_diffL_effect_gammaM}
\end{figure}

To investigate the effect of the dephasing rate $\gamma$ on particle loss, we again consider the total particle loss rate, $dN(t)/dt$, obtained by differentiating the total particle number $N(t)=\sum_i C_{i,i}(t)$ with respect to time. As before, using Eq.~\eqref{EOM_corr_wd}, we obtain
\begin{align}
  \frac{dN}{dt}
  = \sum_i \dot{C}_{i+(i-1)L}(t)
  = \sum_i [MC]_{i+(i-1)L}.
  \label{loss_rate_wd}
\end{align}
The total particle loss rate as a function of time is shown in Fig.~\ref{fig:sink_diffL_effect_gammaM}.

%%%%%%%%%%%%%%%%%%%%%%%%%%%%%%%%%%%%%%%%
%%%%%%%%%%%%%%%%%%%%%%%%%%%%%%%%%%%%%%%%

\begin{figure}[t!]
\centering
\includegraphics[width=0.8\linewidth]{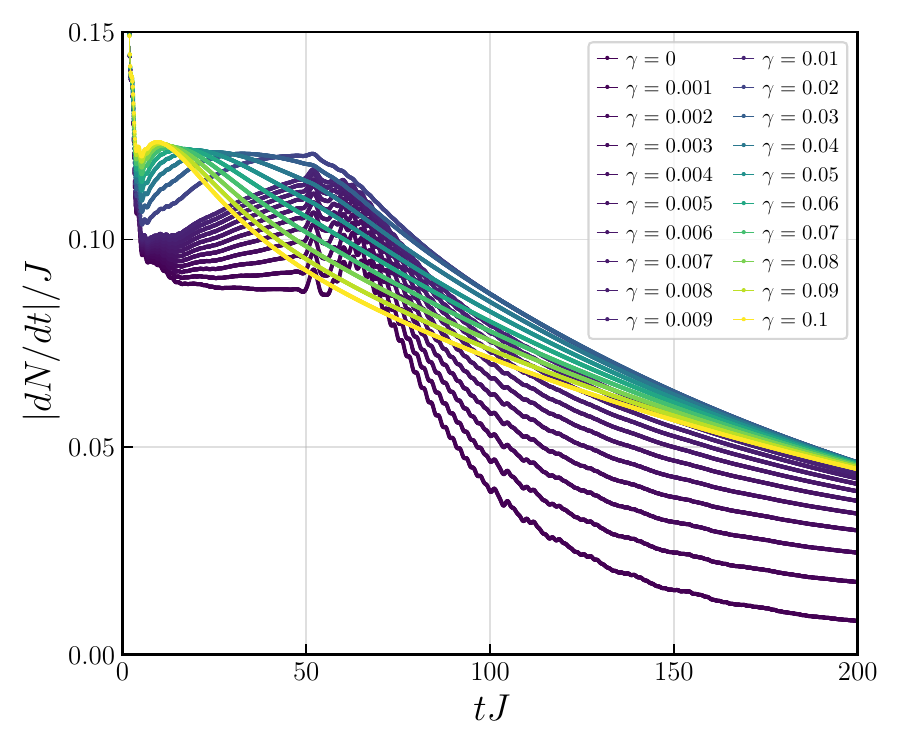}
\caption{Total particle loss rate, $|dN/dt|/J$ [Eq.~\eqref{loss_rate_wd}], as a function of time $tJ$ for a system of size $L=101$ at different dephasing rates $\gamma$. As $\gamma$ increases, the dynamics undergoes a smooth crossover from the three-regime behavior characteristic of coherent transport to the two-regime behavior associated with diffusive transport. The parameters are $J=\Gamma_L=1$, and the initial filling is $N(0)/L=0.25$.} 
\label{fig:effect_gammaM}
\end{figure}
%%%%%%%%%%%%%%%%%%%%%%%%%%%%%%%%%%%%%%%%

Comparing Fig.~\ref{fig:sink_rate_wod} with Fig.~\ref{fig:sink_diffL_effect_gammaM}, we find that dephasing suppresses the system-size dependence of the intermediate regime. Consequently, the loss-rate curves for different system sizes collapse onto a single curve, and the three dynamical regimes observed in the coherent case are reduced to two.

To further elucidate the effect of dephasing, we plot the particle loss rate as a function of time for several values of the dephasing rate $\gamma$. As shown in Fig.~\ref{fig:effect_gammaM}, increasing $\gamma$ gradually suppresses the intermediate ballistic plateau. Consequently, this plateau shrinks and eventually disappears, leaving only two distinct dynamical regimes.

%%%%%%%%%%%%%%%%%%%%%%%%%%%%%%%%%%%%%%%%%%%%%%%
\begin{figure}[t!]
\centering
    \includegraphics[width=0.85\columnwidth]{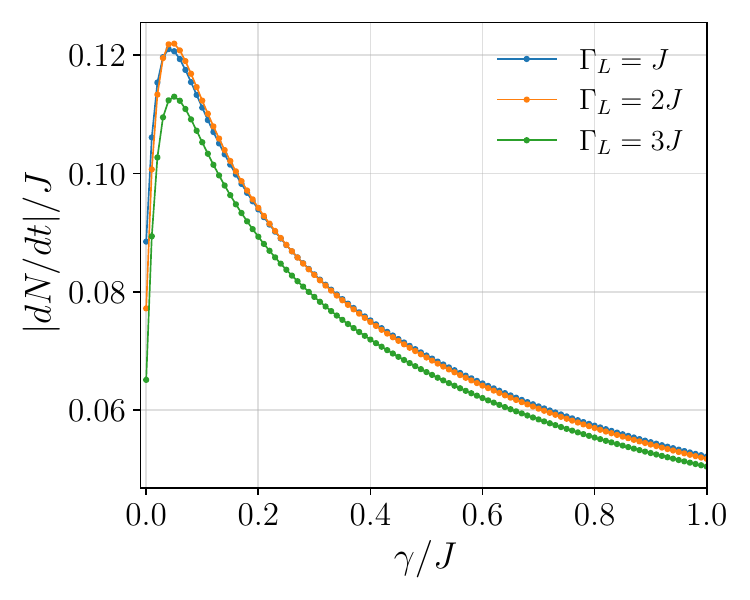}
    \caption{Particle loss rate, $|dN/dt|/J$, measured in the second dynamical regime of Fig.~\ref{fig:sink_rate_wod}, as a function of the dephasing rate $\gamma/J$ for different sink strengths $\Gamma_L$. For weak dephasing, the loss rate initially increases with $\gamma$, demonstrating the anti-quantum Zeno effect. Beyond a threshold dephasing strength, the loss rate decreases, signaling the onset of the quantum Zeno effect. The system size is $L=101$, and $J=1$ for all curves.}
    \label{fig:AQZE}
\end{figure}
%%%%%%%%%%%%%%%%%%%%%%%%%%%%%%%%%%%%%%%%%%%%%%%

\subsection{Quantum Zeno Effect}
Extracting information from a quantum system necessarily involves measurement, which is not without consequence: the act of measurement inevitably perturbs the system and alters its state. Remarkably, performing measurements very frequently to check whether the system remains in its initial state can inhibit its transition to other states, a phenomenon known as the quantum Zeno effect (QZE)~\cite{misra1977zeno}. This effect follows directly from the Schrödinger equation together with the measurement postulate, which implies that, at short times, the survival probability decays quadratically with time~\cite{home1997conceptual}. The QZE has also been observed experimentally~\cite{itano1990quantum,syassen2008strong}. Considerable effort has therefore been devoted to understanding the effect without invoking the measurement postulate, which assumes an instantaneous collapse of the wavefunction, even though real measuring devices require a finite time to register an outcome.

In the context of open quantum systems, measurements can be effectively viewed as continuous interactions with an external environment or measuring apparatus~\cite{Brun2000,Clark2001,Gambetta2008,facchi2002quantum,Filippone2024}. The connection between continuous measurement and the Lindbladian formalism is presented in the Supplemental Material. The manifestation of the QZE in such systems depends on the underlying microscopic dynamics and is typically characterized by the suppression of system dynamics due to coupling with a measuring device~\cite{Krapivsky2019}. In our case, where the measuring device acts as a dephasing probe, this behavior is reflected in the reduction of relevant observables, such as the particle gain or loss rate, with increasing $\gamma$.

For a localized sink, we observe signatures of both the anti-quantum Zeno effect (AQZE)~\cite{Kofman2000} and the QZE, as illustrated in Fig.~\ref{fig:AQZE}, where the particle loss rate is plotted as a function of $\gamma/J$. For small dephasing rates, the particle loss rate initially increases because dephasing suppresses the destructive interference that inhibits particles from reaching the sink, a hallmark of the AQZE~\cite{Kofman2000}. As $\gamma$ is increased further, however, strong dephasing induces Zeno localization, effectively hindering particle transport toward the sink. Consequently, the particle loss rate decreases, signaling the onset of the QZE. The crossover from the AQZE to the QZE depends on the initial particle density $n_0$, as demonstrated in Figs.~\ref{fig:QZE_sink_1} and \ref{fig:QZE_sink_2} of the Supplemental Material.

\subsection{Density profile in the strong-dephasing limit}
In the case of a localized source, the analytical solution for the density profile in the strong-dephasing limit is known from Ref.~\cite{Ray2026}. Following a similar approach, we derive the density profile for the localized sink case in the strong-dephasing limit.

The two-point correlation functions evolve according to
\begin{align}
    \frac{d}{dt}C_{i,j \neq i} &= -iJ\big(C_{i-1,j} + C_{i+1,j} - C_{i,j+1} - C_{i,j-1}\big)  \nonumber \\
&\quad - \Gamma_L(\delta_{i,0} + \delta_{j,0})\,C_{i,j} - 2\gamma C_{i,j}, \label{sink_wd_off}
\end{align}
\begin{align}
    \frac{d}{dt}C_{i,i} &= -iJ\big(C_{i-1,i} + C_{i+1,i} - C_{i,i+1} - C_{i,i-1}\big)  \nonumber \\
&\quad - 2\Gamma_L \delta_{i,0}C_{i,i}. \label{sink_wd_diagonal}
\end{align}
The strong-dephasing limit, also known as the Zeno limit, corresponds to $\gamma \gg J,\Gamma_L$. In this regime, we employ the adiabatic approximation, $\dot{C}_{i,j} \ll \gamma C_{i,j}$, which is justified by the large separation of timescales introduced by the strong dephasing rate. Physically, this approximation implies that coherences relax much faster than populations evolve. Setting the left-hand side of Eq.~\eqref{sink_wd_off} to zero under this approximation, the off-diagonal elements satisfy
\begin{align}
    C_{i,j \neq i} =
    \frac{-iJ\left(C_{i+1,j}+C_{i-1,j}-C_{i,j+1}-C_{i,j-1}\right)}
    {2\gamma+\Gamma_L(\delta_{i,0}+\delta_{j,0})}.
    \label{evolution-off}
\end{align}
Using Eq.~\eqref{evolution-off}, we eliminate the nearest-neighbor coherences from Eq.~\eqref{sink_wd_diagonal}. Furthermore, neglecting higher-order coherences (i.e., $C_{i,j}$ with $|i-j|>1$), we obtain a closed equation of motion for the diagonal elements,
\begin{align}
    \frac{dC_{\mathrm{diag}}}{dt}
    =
    \mathbb{W}C_{\mathrm{diag}},
    \label{diag-mat}
\end{align}
where $C_{\mathrm{diag}}(t)$ is the vector of local densities whose $i$th component is $n_i(t)$. The matrix $\mathbb{W}$ is given by
\[
\resizebox{\columnwidth}{!}{$
\mathbb{W} =
\begin{pmatrix}
-2\omega_1 & \omega_1 & 0 & 0 & 0 & 0 & \cdots & \omega_1 \\
\omega_1 & -2\omega_1 & \omega_1 & 0 & 0 & 0 & \cdots & 0 \\
0 & \omega_1 & -2\omega_1 & \omega_1 & 0 & 0 & \cdots & 0 \\
\vdots & \vdots & \vdots & \vdots & \vdots & \vdots & \vdots & \vdots\\
0 & \cdots & \omega_1 & -(\omega_1+\omega_2) & \omega_2 & \cdots & \cdots & 0\\
0 & \cdots & 0 & \omega_2 & -2\omega_2-2\Gamma_L & \omega_2 & \cdots & 0\\
0 & \cdots & 0 & 0 & \omega_2 & -(\omega_1+\omega_2) & \omega_1 & 0\\
\vdots & \cdots & \cdots & 0\\
\omega_1 & 0 & 0 & \cdots & 0 & 0 & \omega_1 & -2\omega_1
\end{pmatrix}
$}
\]
with
\begin{align}
    \omega_1=\frac{J^2}{\gamma},
    \qquad
    \omega_2=\frac{2J^2}{2\gamma+\Gamma_L}.
    \label{omega}
\end{align}
The row containing the term $-2\Gamma_L$ corresponds to the lattice site coupled to the localized sink.

\begin{figure}[t!]
\centering
    \includegraphics[width=0.85\linewidth]{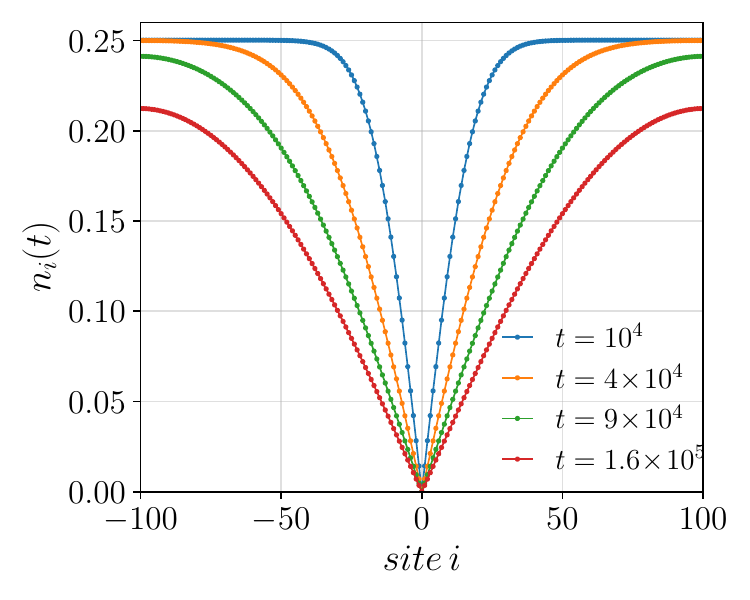}
    \caption{Local particle density profile $n_i(t)$ as a function of the lattice site $i$ in the strong-dephasing limit, obtained from Eq.~\eqref{formal_sol} within the adiabatic approximation. The results are shown for a system of size $L=201$, with the matrix $\mathbb{W}$ and parameters $\omega_1$ and $\omega_2$ given by Eq.~\eqref{omega}. The parameters are $J=\Gamma_L=1$ and $\gamma=100J$.}
    \label{fig:density_dephase}
\end{figure}

The formal solution of Eq.~\eqref{diag-mat} is
\begin{align}
    C_{\mathrm{diag}}(t)
    =
    e^{\mathbb{W}t}C_{\mathrm{diag}}(0),
    \label{formal_sol}
\end{align}
where the $i$th component of $C_{\mathrm{diag}}(t)$ gives the local density $n_i(t)$. The resulting density profiles at different times are shown in Fig.~\ref{fig:density_dephase}.

\begin{figure}[t!]
    \includegraphics[width=0.85\linewidth]{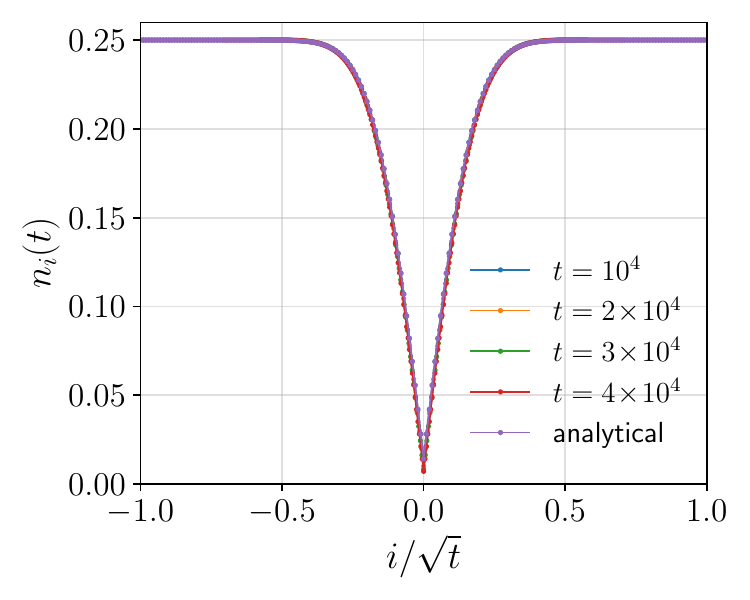}
    \caption{Density profile $n_i(t)$ plotted against the scaled coordinate $i/\sqrt{t}$ in the strong-dephasing limit within the adiabatic approximation. The numerical results are obtained from Eq.~\eqref{formal_sol} for a system of size $L=201$, with the matrix $\mathbb{W}$ given by Eq.~\eqref{adia_Wmat} and the sink strength $\Gamma_L$ specified by Eq.~\eqref{omega1}. The solid line represents the analytical prediction from Eq.~\eqref{42}. The parameters are $J=\Gamma_L=1$, $\gamma=100J$, and the initial filling is $N(0)/L=0.25$.}
    \label{fig:scaled_adiabatic_density}
\end{figure}

The density profile obeys the diffusive scaling form $n_i(t)=f(i/\sqrt{t})$, as evidenced by the collapse of the profiles at different times onto a single curve when plotted against $i/\sqrt{t}$ (see Fig.~\ref{fig:scaled_adiabatic_density}). This confirms that the dynamics is diffusive in the strong-dephasing limit.

We can further simplify the evolution matrix in the strong-dephasing limit ($\gamma \gg J,\Gamma_L$). In this regime,
\[
2\gamma+\Gamma_L \simeq 2\gamma,
\]
so that $\omega_2 \simeq \omega_1$. In addition, assuming
\begin{align}
\Gamma_L = \frac{J^2}{\gamma},
\label{omega1}
\end{align}
the matrix $\mathbb{W}$ reduces to
\begin{align}
\mathbb{W} &= \omega_1
\begin{pmatrix}
-2 & 1 & 0 & 0 & 0 & 0 & \cdots & 1 \\
1 & -2 & 1 & 0 & 0 & 0 & \cdots & 0 \\
0 & 1 & -2 & 1 & 0 & 0 & \cdots & 0 \\
\vdots & \vdots & \vdots & \vdots & \vdots & \vdots & \vdots & \vdots\\
0 & \cdots & 1 & -2 & 1 & \cdots & \cdots & 0\\
0 & \cdots & 0 & 1 & -4 & 1 & \cdots & 0 \\
0 & \cdots & 0 & 0 & 1 & -2 & 1 & 0\\
\vdots & \cdots & \cdots & 0 & 0 & 1 & -2 & 1\\
1 & 0 & 0 & \cdots & 0 & 0 & 1 & -2
\end{pmatrix}.
\label{adia_Wmat}
\end{align}

From the structure of the evolution matrix, we observe that it is a circulant matrix with a single defect (impurity) at the origin. In the continuum limit, Eq.~\eqref{diag-mat} reduces to the diffusion equation with a localized sink,
\begin{align}
    \frac{\partial n(x,t)}{\partial t}
    =
    \omega_1
    \left[
    \frac{\partial^2 n(x,t)}{\partial x^2}
    -
    2\delta(x-x_0)n(x,t)
    \right].
\end{align}
Without loss of generality, we choose the defect to be located at the origin, $x_0=0$, so that
\begin{align}
    \frac{\partial n(x,t)}{\partial t}
    =
    \omega_1\frac{\partial^2 n(x,t)}{\partial x^2}
    -
    2\omega_1\delta(x)n(x,t).
    \label{continuum limit}
\end{align}
Equation~\eqref{continuum limit}, together with the initial condition $n(x,0)=n_0(x)$ and boundary condition $n(x,t)\rightarrow0$ as $|x|\rightarrow\infty$, can be solved exactly using the Laplace--Fourier transform (see Appendix~\ref{A1} for the derivation).

The resulting density profile is obtained as
\begin{align}
n(x,t)
&=
\frac{k_F}{\pi}\,
\operatorname{erf}\!\left(\frac{|x|}{2\sqrt{\omega_1t}}\right)
\notag\\
&\quad+
\frac{k_F}{\pi}\,
e^{\,|x|+\omega_1t}
\operatorname{erfc}\!\left(
\frac{|x|+2\omega_1t}{2\sqrt{\omega_1t}}
\right),
\label{42}
\end{align}
which constitutes the exact solution to Eq.~\eqref{continuum limit}.

The analytical solution, Eq.~\eqref{42}, exhibits the diffusive scaling form $n(x,t)=f(x/\sqrt{t})$ and is in excellent agreement with the numerical solution of Eq.~\eqref{diag-mat}, as evidenced by the scaling collapse shown in Fig.~\ref{fig:scaled_adiabatic_density}.

\section{Localized source}\label{source}
Here, we consider the case of a localized source in the absence of a sink term.

\subsection{Localized source without dephasing}
We first consider a one-dimensional periodic lattice coupled to a localized source of fermions at the origin. The dynamics of the open quantum system is governed by the Lindblad master equation
\begin{align}
\partial_t \rho
=
-i[H,\rho]
+
\mathcal{L}_{\mathrm{source}}\rho,
\label{source_wod}
\end{align}
where the source Lindblad superoperator is
\begin{align}
\mathcal{L}_{\mathrm{source}}\rho
=
\Gamma_G
\left(
2a_0^{\dagger}\rho a_0
-
\{a_0a_0^{\dagger},\rho\}
\right),
\label{source_diss1}
\end{align}
with $\Gamma_G$ denoting the particle injection rate. We assume that the lattice is initially empty. At $t=0$, the localized source is switched on, injecting fermions into the lattice, which subsequently spread through coherent hopping governed by the Hamiltonian in Eq.~\eqref{PBC_H}.

The average total number of fermions and the local density can be obtained from the two-point correlation function, defined in Eq.~\eqref{def_corr}. Using the Lindblad master equation~\eqref{source_wod}, together with the Hamiltonian~\eqref{PBC_H}, the source dissipator~\eqref{source_diss1}, and the fermionic anticommutation relations, we obtain the equation of motion
\begin{align}
 \frac{d}{dt}C_{i,j}
 &= -iJ\big(C_{i-1,j}+C_{i+1,j}-C_{i,j+1}-C_{i,j-1}\big)
 \nonumber\\
&\quad
-\Gamma_G(\delta_{i,0}+\delta_{j,0})\,C_{i,j}
+2\Gamma_G\,\delta_{i,0}\delta_{j,0}.
\label{non_homo_source_wod}
\end{align}
Notice that, as in the sink case, Eq.~\eqref{non_homo_source_wod} can be recast in the form
\begin{align}
    \frac{dC}{dt} = i[W,C] + \{L,C\} + P 
    \label{mat_corr_source_wod}
\end{align}
where the matrix $W$ has the same form as in the sink case, while $L$ and $P$ are given by
\begin{align}
L_{i,j} &= -\Gamma_G \delta_{i,0}\delta_{i,j}, \\
P_{i,j} &= 2\Gamma_G \delta_{i,0}\delta_{j,0}.
\label{L_P_form}
\end{align}
The matrix $P$ is the only term absent in the corresponding sink problem. Given $C(0)=0$, the formal solution of Eq.~\eqref{mat_corr_source_wod} is
\begin{align}
    C(t)=\int_{0}^{t} d\tau \,
    e^{G(t-\tau)} P\, e^{G^{\dagger}(t-\tau)}
    \label{source_sol}
\end{align}
where
\begin{align}
    G=iW+L.
\end{align}

\begin{figure}[t!]
    \includegraphics[width=0.85\linewidth]{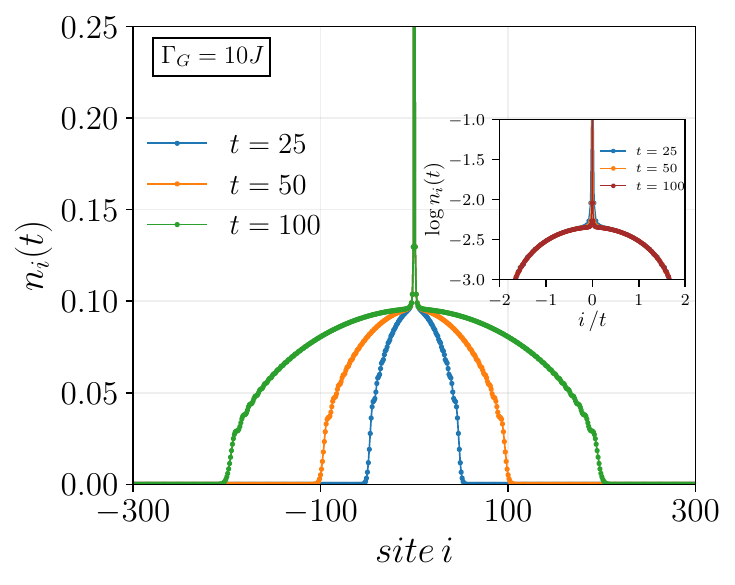}
     \caption{Local particle density profiles $n_i(t)$ [Eq.~\eqref{soure_density_wod}] as a function of the lattice site $i$ at different times for a system of size $L=601$, with a localized source at the origin and injection rate $\Gamma_G$. The inset shows the scaled density $\ln[n_i(t)]$ plotted against the scaled coordinate $i/t$, demonstrating the ballistic spreading of the density perturbation. $J=1$ for all curves.}
    \label{fig:source_density_wod}
\end{figure}

\subsubsection{Local density profile}
Following the same procedure as in the sink case, we diagonalize the matrices $G$ and $G^{\dagger}$ as
\[
G=V\lambda^1 V^{-1},\qquad
G^{\dagger}=U\lambda^2 U^{-1},
\]
where $\lambda^1=\mathrm{diag}(\lambda_1^1,\lambda_2^1,\ldots)$ and $\lambda^2=\mathrm{diag}(\lambda_1^2,\lambda_2^2,\ldots)$ are the corresponding diagonal matrices of eigenvalues. The local density is then given by
\begin{align}
    n_i(t)
    =
    \sum_{\alpha,\beta}
    \left[
    \frac{e^{(\lambda_{\alpha}^{1}+\lambda_{\beta}^{2})t}-1}
    {\lambda_{\alpha}^{1}+\lambda_{\beta}^{2}}
    \right]
    V_{i\alpha}\,
    X_{\alpha\beta}\,
    U^{-1}_{\beta i},
    \label{soure_density_wod}
\end{align}
where $X=V^{-1}PU$.

The local density profiles at different times are shown in Fig.~\ref{fig:source_density_wod}. As the source strength $\Gamma_G$ is increased relative to the hopping amplitude $J$, the particle density at the origin increases, while the spatial spreading of particles is progressively suppressed. This behavior can be observed by comparing Fig.~\ref{fig:source_density_wod} with Fig.~\ref{fig:sup_source_wod} in the Supplementary Material. The density profiles exhibit ballistic propagation, as evidenced by the scaling collapse shown in the inset of Fig.~\ref{fig:source_density_wod}. Similar ballistic transport has previously been reported for systems with open boundary conditions in Ref.~\cite{Trivedi2023}.

\subsubsection{Different time regimes}
\begin{figure}[t!]
    \centering
    \includegraphics[width=0.85\linewidth]{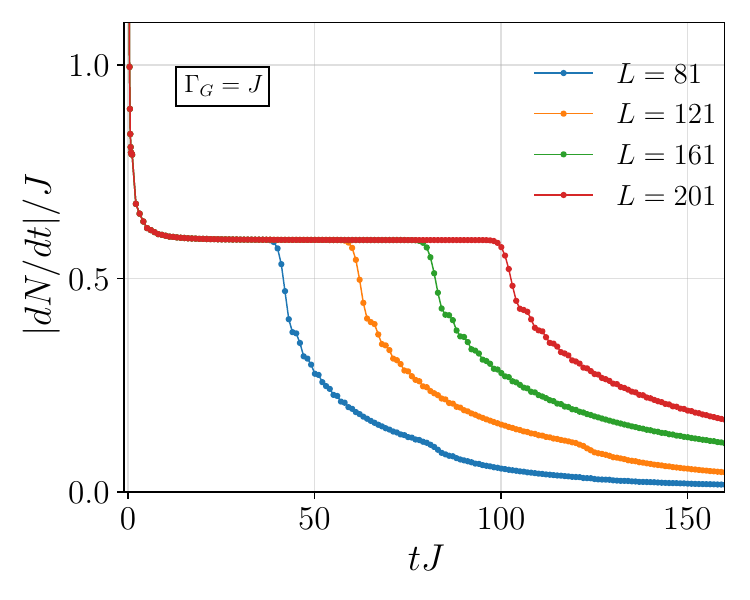}
    \caption{Total particle growth rate, $|dN/dt|/J$ [Eq.~\eqref{growth_rate}], as a function of time for different system sizes $L$, with a localized source at the origin injecting particles at rate $\Gamma_G$. The parameters are $\Gamma_G=J$ and $J=1$.}
    \label{fig:source_rate_wod}
\end{figure}

Next, we consider the total particle growth rate, $dN(t)/dt$. Following the same procedure as in the sink case and using Eq.~\eqref{source_sol}, we obtain
\begin{align}
    \frac{dN}{dt}
    =
    \sum_i \left[P+GC+CG^{\dagger}\right]_{ii}(t).
    \label{growth_rate}
\end{align}
The growth rate, shown in Fig.~\ref{fig:source_rate_wod}, exhibits three distinct regimes: an initial transient, an intermediate ballistic regime characterized by a constant growth rate, and a long-time finite-size regime in which the growth rate decays to zero. In the thermodynamic limit, the ballistic regime persists indefinitely. A comparison of the loss rate in the sink case (Fig.~\ref{fig:sink_rate_wod}) with the growth rate in the source case (Fig.~\ref{fig:source_rate_wod}) highlights the underlying symmetry between particle absorption and injection in the coherent limit.

\begin{figure}[t!]
    \centering
    \includegraphics[width=0.85\linewidth]{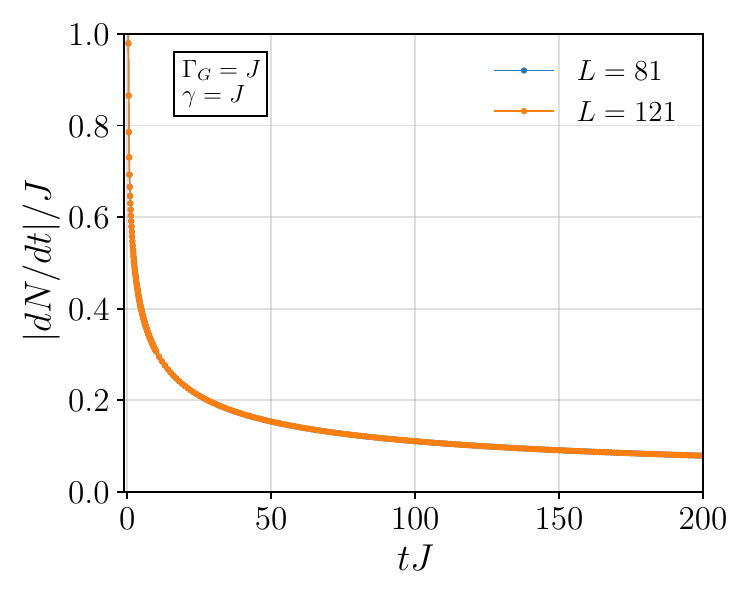}
    \caption{Total particle growth rate, $|dN/dt|/J$ [Eq.~\eqref{growth_rate_wd}], as a function of time for different system sizes $L$, with a localized source at the origin injecting particles at rate $\Gamma_G$ and local dephasing of strength $\gamma$ acting on every lattice site. In the presence of dephasing, ballistic transport is suppressed, reducing the three dynamical regimes observed in the coherent case to two. The parameters are $J=\Gamma_G=\gamma=1$.}
    \label{fig:diffL_effect_gammaM}
\end{figure}

\subsection{Localized source with dephasing}
In the presence of local dephasing, the dynamics is governed by the Lindblad master equation
\begin{align}
\partial_t \rho
=
-i[H,\rho]
+\mathcal{L}_{\mathrm{source}}\rho
+\mathcal{L}_{\mathrm{deph}}\rho,
\label{lind_source_dephase}
\end{align}
where the source dissipator $\mathcal{L}_{\mathrm{source}}\rho$ is given by Eq.~\eqref{source_diss1}, and the dephasing dissipator is
\begin{align}
    \mathcal{L}_{\mathrm{deph}}\rho
    =
    \gamma\sum_i
    \left(
    2\hat n_i\rho\hat n_i
    -
    \{\hat n_i^2,\rho\}
    \right),
    \label{dephase_ope}
\end{align}
with $\gamma$ denoting the dephasing rate. Using Eqs.~\eqref{PBC_H}, \eqref{source_diss1}, and \eqref{dephase_ope}, we obtain the equation of motion for the two-point correlation function,
\begin{align}
 \frac{d}{dt}C_{i,j}
 &= -iJ\big(C_{i-1,j}+C_{i+1,j}-C_{i,j+1}-C_{i,j-1}\big)
 \nonumber\\
 &\quad
 -\Gamma_G(\delta_{i,0}+\delta_{j,0})C_{i,j}
 +2\gamma(\delta_{i,j}-1)C_{i,j}
 \nonumber\\
 &\quad
 +2\Gamma_G\delta_{i,0}\delta_{j,0}.
 \label{EOM_source_depha}
\end{align}
It is evident from Eq.~\eqref{EOM_source_depha} that the inclusion of the dephasing operator suppresses the off-diagonal elements of the correlation matrix, leading to the decay of quantum coherences while leaving the diagonal populations unaffected.

Following the same vectorization procedure as in the sink case, Eq.~\eqref{EOM_source_depha} can be written in the compact form
\begin{align}
    \dot{C}=MC+P,
    \label{source_dephase_vec}
\end{align}
where $M$ is the $L^2\times L^2$ coefficient matrix corresponding to the homogeneous terms in Eq.~\eqref{EOM_source_depha}, and $P$ is the $L^2$-dimensional source vector arising from the inhomogeneous term $2\Gamma_G\delta_{i,0}\delta_{j,0}$. The formal solution is
\begin{align}
    C(t)
    =
    \int_{0}^{t} d\tau\, e^{M(t-\tau)}P.
    \label{evo_vec_corr}
\end{align}

\subsubsection{Local density profile}
The local density profile at time $t$ can be obtained by solving Eq.~\eqref{evo_vec_corr}. Diagonalizing the matrix $M$ as $M = U \Lambda U^{-1}$, the $r$-th component of Eq.~\eqref{evo_vec_corr} is given by

\begin{align}
    C_r(t) = \sum_{\alpha,\beta} \left[\frac{e^{\lambda^M_{\alpha} t}-1}{\lambda^M_{\alpha}}\right] U_{r\alpha} \, U_{\alpha \beta}^{-1} P_{\beta}.
    \label{sol_source_dephase}
\end{align}
This expression enables the calculation of both the local density and two-point spatial correlations. The local density corresponds to the diagonal elements of the correlation matrix,
\begin{align}
    n_i(t) = C_{i+(i-1)L}(t).
\end{align}

\begin{figure}[t!]
    \centering
    \includegraphics[width=0.85\linewidth]{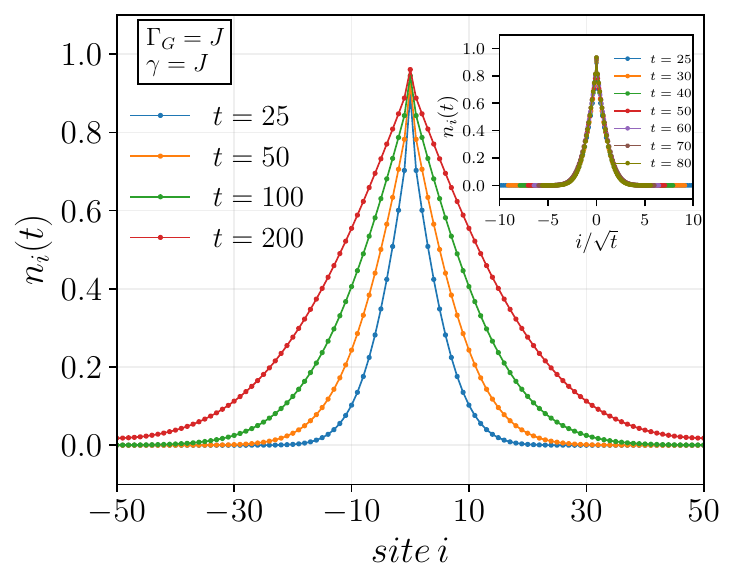}
    \caption{Local particle density profiles $n_i(t)$ [Eq.~\eqref{sol_source_dephase}] as a function of the lattice site $i$ at different times for a system of size $L=101$, with a localized source at the origin injecting particles at rate $\Gamma_G$ and local dephasing of strength $\gamma$ acting on every lattice site. The inset shows the density profiles plotted against the scaled coordinate $i/\sqrt{t}$, demonstrating the diffusive spreading of the density perturbation. The parameters are $J=\Gamma_G=\gamma=1$.}
    \label{fig:source_density_wd}
\end{figure}

The local density profile in the presence of dephasing is shown in Fig.~\ref{fig:source_density_wd} at different times. As illustrated in the inset, the density profile spreads diffusively, in contrast to the ballistic behavior observed without dephasing (Fig.~\ref{fig:source_density_wod}). As shown in Ref.~\cite{Ray2026} for an open-geometry setup, the system undergoes a crossover from ballistic to diffusive transport at a timescale determined by the interplay among $\Gamma_G$, $\gamma$, and $J$. In Fig.~\ref{fig:source_density_wd}, parameters are chosen such that the diffusive regime is reached early in the dynamics. Additional plots illustrating different values of source and dephasing strengths alongside their scaling forms are provided in Figs.~\ref{fig:sup_source_wd_fig2} and \ref{fig:sup_scaled_source_wd} of the Supplemental Material.

\subsubsection{Different time regimes}
Similar to the sink case, introducing dephasing eliminates the three distinct dynamical regimes present in the coherent limit, reducing them to two (Fig.~\ref{fig:diffL_effect_gammaM}).

To analyze the effect of $\gamma$ on the particle growth rate, we evaluate $dN(t)/dt$ by taking the time derivative of the total particle number $N(t) = \sum_i C_{i,i}(t)$. From Eq.~\eqref{source_dephase_vec}, we obtain
\begin{align}
    \frac{dN}{dt} = \sum_i \dot{C}_{i+(i-1)L}(t) = \sum_i[MC + P]_{i+(i-1)L}.
    \label{growth_rate_wd}
\end{align}
The total particle growth rate as a function of time is shown in Fig.~\ref{fig:diffL_effect_gammaM}.

\begin{figure}[t!]
    \centering
    \includegraphics[width=0.85\columnwidth]{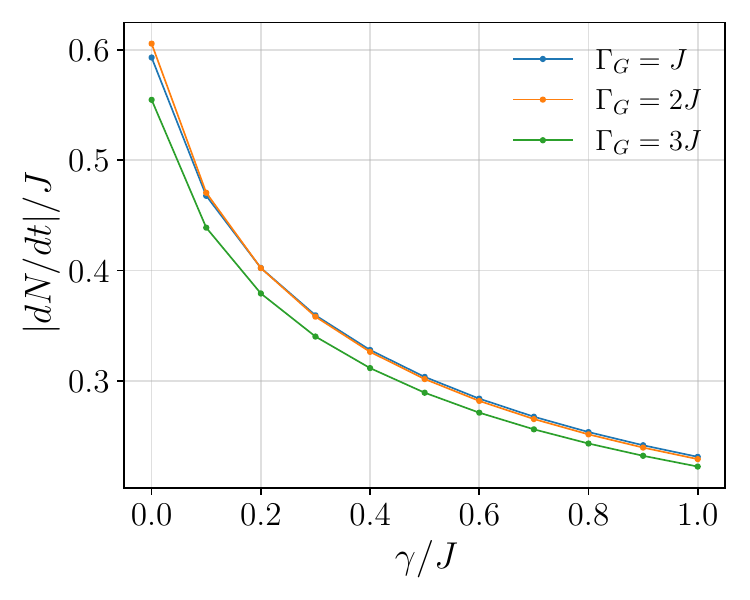}
    \caption{Total particle growth rate, $|dN/dt|/J$ [Eq.~\eqref{growth_rate_wd}], as a function of the dephasing strength $\gamma/J$ for different source strengths $\Gamma_G$. The growth rate is evaluated at times corresponding to the ballistic growth regime in the absence of dephasing (the second regime in Fig.~\ref{fig:source_rate_wod}). For all curves, $J=1$. The suppression of the growth rate with increasing $\gamma$ provides a clear signature of the quantum Zeno effect, where system dynamics are progressively inhibited as the effective measurement rate increases.}
    \label{fig:QZE}
\end{figure}

While the quantum Zeno effect (QZE) in the presence of localized loss has been studied extensively, much less is understood in the case of localized gain. In Fig.~\ref{fig:QZE}, we plot the particle gain rate as a function of $\gamma/J$. The suppression of the gain rate with increasing $\gamma/J$ provides a clear signature of the QZE.

\section{Localized sink and source}\label{sink_source}

We now investigate the dynamics of a one-dimensional lattice in the presence of both a localized source and a localized sink separated by a distance $\Delta$. We consider both coherent dynamics and dynamics under local dephasing for three different initial conditions.

In all cases, the finite separation $\Delta$ gives rise to nonlocal secondary density peaks located at integer multiples of $\Delta$ at sufficiently long times. These secondary peaks are significantly weaker than the primary density peaks and constitute a purely quantum effect arising from coherent interference. They disappear in the presence of dephasing, demonstrating that they are a direct consequence of quantum coherence.

\subsection{Localized sink and source without dephasing}
The dynamics in the presence of a localized source at the origin and a localized sink at site $\Delta$ is governed by the Lindblad master equation
\begin{align}
\partial_t \rho
=
-i[H,\rho]
+\mathcal{L}_{\mathrm{sink}}\rho
+\mathcal{L}_{\mathrm{source}}\rho,
\label{lind_sink_source_wod}
\end{align}
where the sink dissipator is
\begin{align}
\mathcal{L}_{\mathrm{sink}}\rho
=
\Gamma_L\left(
2a_{\Delta}\rho a_{\Delta}^{\dagger}
-
\{a_{\Delta}^{\dagger}a_{\Delta},\rho\}
\right),
\label{sink_diss1}
\end{align}
and the source dissipator $\mathcal{L}_{\mathrm{source}}\rho$ is given by Eq.~\eqref{source_diss1}.

The particle density profile and average total number of fermions are obtained from the two-point correlation function $C_{i,j}(t)$. Using the master equation~\eqref{lind_sink_source_wod} together with the Hamiltonian~\eqref{PBC_H} and dissipators~\eqref{sink_diss1} and \eqref{source_diss1}, we obtain
\begin{align}
 \frac{d}{dt}C_{i,j}
 &= -iJ\big(C_{i-1,j}+C_{i+1,j}-C_{i,j+1}-C_{i,j-1}\big)
 \nonumber\\
 &\quad
 -\Gamma_G(\delta_{i,0}+\delta_{j,0})C_{i,j}
 -\Gamma_L(\delta_{i,\Delta}+\delta_{j,\Delta})C_{i,j}
 \nonumber\\
 &\quad
 +2\Gamma_G\delta_{i,0}\delta_{j,0}.
 \label{non_homo_sink_source_wod}
\end{align}
Following the same procedure as in the source-only case, Eq.~\eqref{non_homo_sink_source_wod} can be expressed as
\begin{align}
\frac{dC}{dt}=i[W,C]+\{L,C\}+P,
\label{mat_corr_sink_source_wod}
\end{align}
where $P$ is given by Eq.~\eqref{L_P_form}, and the matrix $L$ is
\begin{align}
L_{i,j}
=
-\left(\Gamma_G\delta_{i,0}
+\Gamma_L\delta_{i,\Delta}\right)\delta_{i,j}.
\label{L_sink_source}
\end{align}
The formal solution is
\begin{align}
C(t)
=
e^{Gt}C(0)e^{G^{\dagger}t}
+
\int_0^t d\tau\,
e^{G(t-\tau)}
P
e^{G^{\dagger}(t-\tau)},
\label{full_corr_sink_source_wod}
\end{align}
where $G=iW+L$. Diagonalizing $G=V\Lambda_1V^{-1}$ and $G^{\dagger}=U\Lambda_2U^{-1}$, the local density is expressed as
\begin{align}
n_i(t)
&=
\sum_{\alpha,\beta}
e^{(\lambda_\alpha^1+\lambda_\beta^2)t}
V_{i\alpha}
Y_{\alpha\beta}
U^{-1}_{\beta i}
\nonumber\\
&\quad
+
\sum_{\alpha,\beta}
\left[
\frac{e^{(\lambda_\alpha^1+\lambda_\beta^2)t}-1}
{\lambda_\alpha^1+\lambda_\beta^2}
\right]
V_{i\alpha}
X_{\alpha\beta}
U^{-1}_{\beta i},
\label{general_source_sink_wod_mat}
\end{align}
where $Y=V^{-1}C(0)U$ and $X=V^{-1}PU$.

We consider three different initial conditions:
\begin{enumerate}
    \item Fermi sea at filling $n_0$,
    \item Completely filled lattice,
    \item Empty lattice.
\end{enumerate}

The latter two initial conditions are \enquote{classical} in the sense that particles occupy localized spatial positions, whereas the first initial condition represents a quantum many-body state with delocalized particles and long-range coherence. We investigate how these initial conditions influence both the transient dynamics and the steady-state properties of the system.

For the Fermi sea, the system is initially prepared at zero temperature with Fermi momentum $k_F=\pi n_0$, where $n_0$ is the initial particle density~\cite{Ultracold}. The corresponding initial correlation matrix is
\begin{align}
C_{m,n}(0)
=
\frac{1}{L}
\sum_{k,k'}
e^{-ikm}
e^{ik'n}
\langle
\tilde{a}^{\dagger}_k(0)
\tilde{a}_{k'}(0)
\rangle.
\label{ini_partial}
\end{align}
For a completely filled lattice,
\begin{align}
C_{i,j}(0)=\delta_{i,j},
\label{ini_full}
\end{align}
whereas for an empty lattice,
\begin{align}
C_{i,j}(0)=0.
\label{ini_empty}
\end{align}

\begin{figure*}[t!]
\centering
\subfloat[]{%
\includegraphics[width=0.32\textwidth]{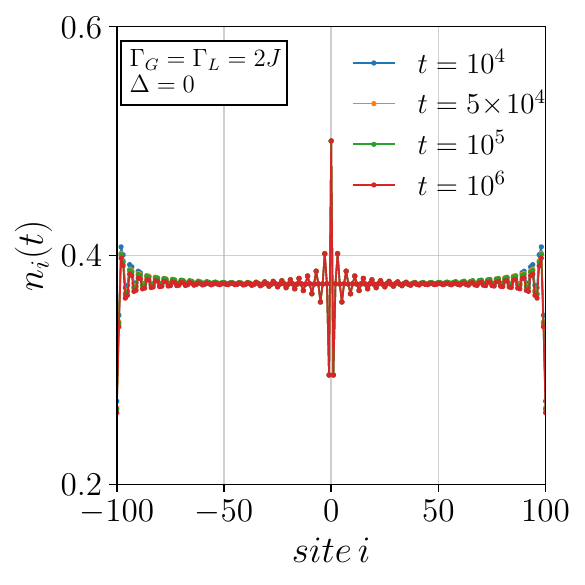}}
\hfill
\subfloat[]{%
\includegraphics[width=0.32\textwidth]{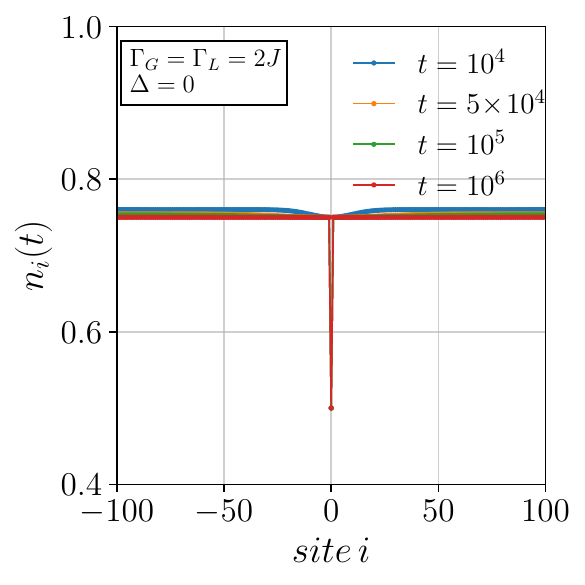}}
\hfill
\subfloat[]{%
\includegraphics[width=0.32\textwidth]{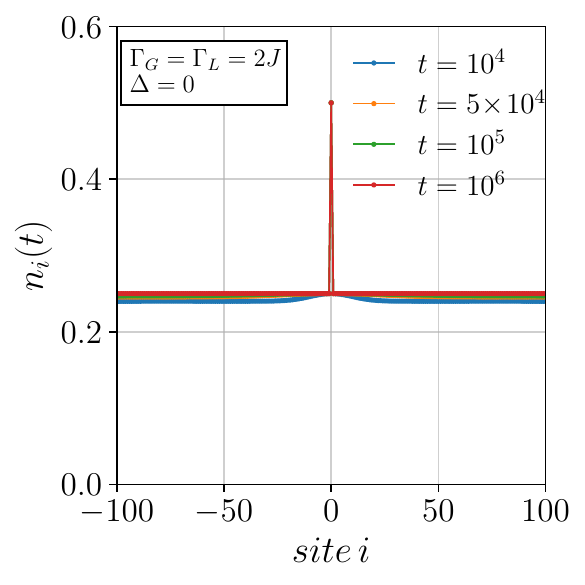}}
\caption{Local particle density profiles $n_i(t)$ [Eq.~\eqref{general_source_sink_wod_mat}] as a function of the lattice site $i$ at sufficiently long times for a system of size $L=201$, with both the localized source and the sink attached at the origin ($\Delta=0$). (a) Initial Fermi sea with filling $N(0)/L=0.25$ [Eq.~\eqref{ini_partial}]. In this case, the system evolves to a nonequilibrium steady state in which the Friedel oscillations persist. (b) Initially fully occupied lattice [Eq.~\eqref{ini_full}]. (c) Initially empty lattice [Eq.~\eqref{ini_empty}]. For the initially full and empty lattices, the density profiles relax to steady states characterized by a single peak at the origin with no additional spatial features. The hopping amplitude is $J=1$ for all curves.}
\label{fig:delta0_wod}
\end{figure*}

The density profiles for the three initial conditions in the $\Delta=0$ case (where both the source and sink reside at the origin) are shown in Fig.~\ref{fig:delta0_wod}. For the Fermi sea initial condition, the system evolves into a nonequilibrium steady state in which Friedel oscillations persist. This is evidenced by the convergence of the density profiles at long times, as shown in Fig.~\ref{fig:delta0_wod}(a). In contrast, for the classical initial conditions (completely filled and empty lattices), the density profiles evolve into a steady state characterized by a single peak at the origin without additional spatial structure [Fig.~\ref{fig:delta0_wod}(b) and (c)].

\begin{figure*}[t!]
\centering
\subfloat[]{%
\includegraphics[width=0.32\textwidth]{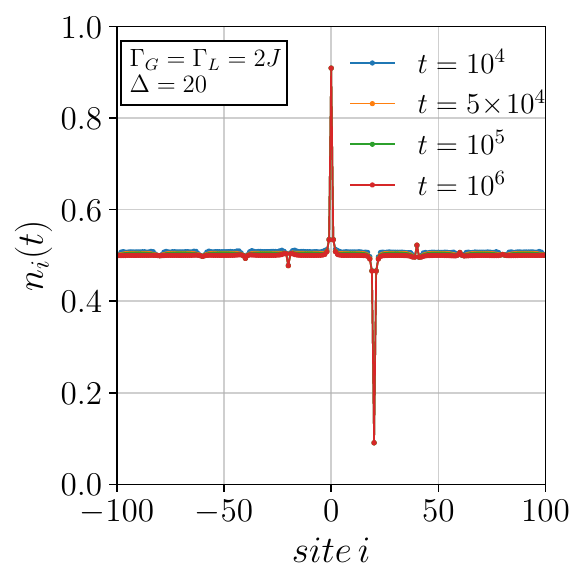}}
\hfill
\subfloat[]{%
\includegraphics[width=0.32\textwidth]{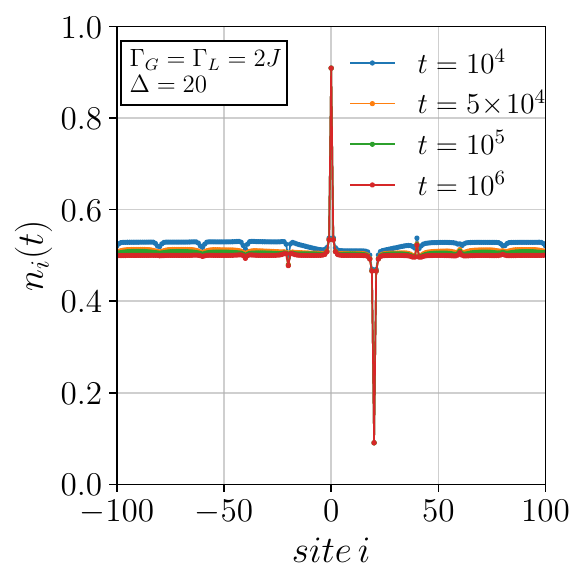}}
\hfill
\subfloat[]{%
\includegraphics[width=0.32\textwidth]{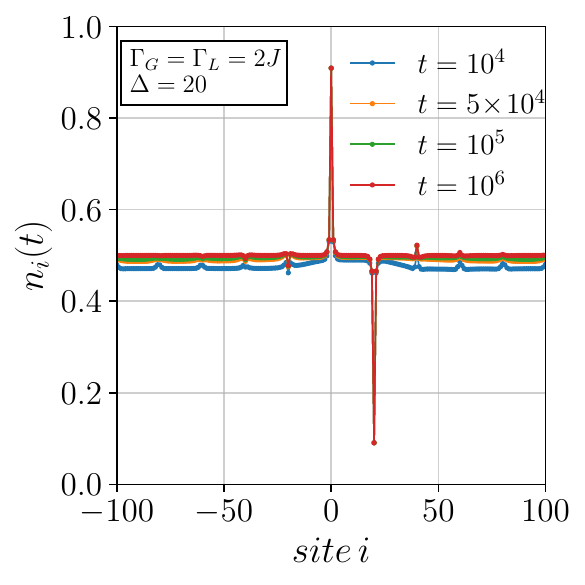}}
\caption{Local particle density profiles $n_i(t)$ [Eq.~\eqref{general_source_sink_wod_mat}] as a function of the lattice site $i$ at sufficiently long times for a system of size $L=201$, with a localized source at the origin and a localized sink located at a distance $\Delta=20$. The three panels correspond to different initial conditions: (a) a partially filled Fermi sea with filling $N(0)/L=0.25$ [Eq.~\eqref{ini_partial}], (b) an initially fully occupied lattice [Eq.~\eqref{ini_full}], and (c) an initially empty lattice [Eq.~\eqref{ini_empty}]. Although the transient dynamics depend on the initial condition, all three cases evolve to the same nonequilibrium steady state. In addition to the primary density peaks at the source and sink, secondary density peaks emerge at positions separated by integer multiples of $\Delta$, providing a clear signature of coherent quantum interference generated during the dynamics. The hopping amplitude is $J=1$ for all curves.}
\label{fig:delta20_wod}
\end{figure*}

For $\Delta\neq 0$, the system evolves into a unique nonequilibrium steady state that is independent of the initial condition, as illustrated in Fig.~\ref{fig:delta20_wod} for $\Delta=20$. In addition to the primary density peaks at the source and sink locations, secondary density peaks emerge at positions separated by integer multiples of $\Delta$. These secondary peaks represent a purely quantum effect arising from coherent interference during the dynamics. Remarkably, they persist in the steady state irrespective of whether the initial state possesses quantum coherence (Fermi sea) or is initially incoherent (completely filled or empty lattice).

\begin{figure*}[t!]
\centering
\subfloat[]{%
\includegraphics[width=0.32\textwidth]{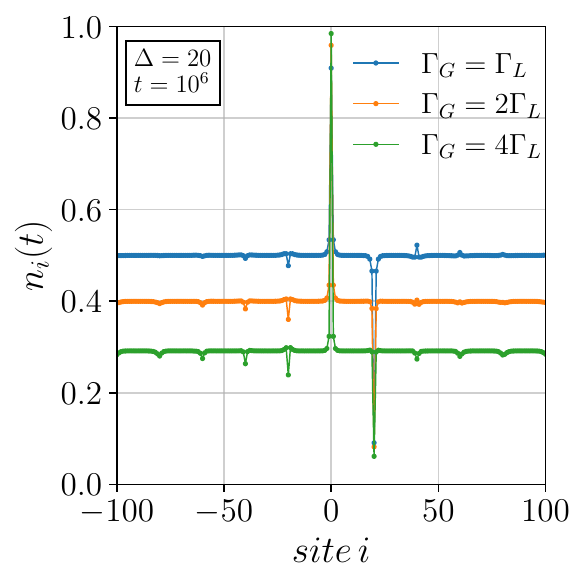}}
\hfill
\subfloat[]{%
\includegraphics[width=0.32\textwidth]{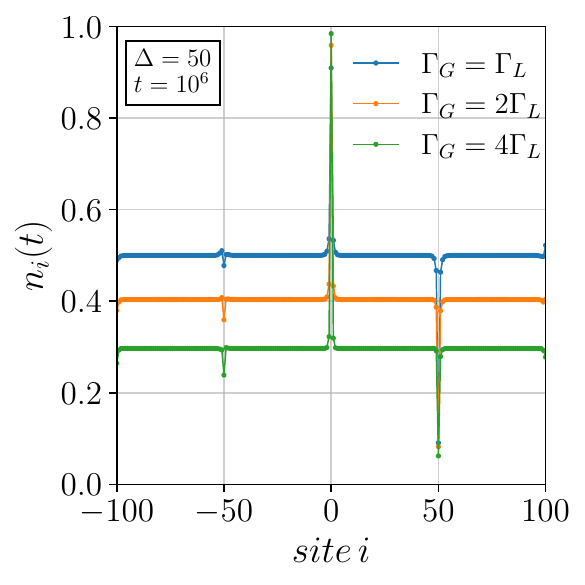}}
\hfill
\subfloat[]{%
\includegraphics[width=0.32\textwidth]{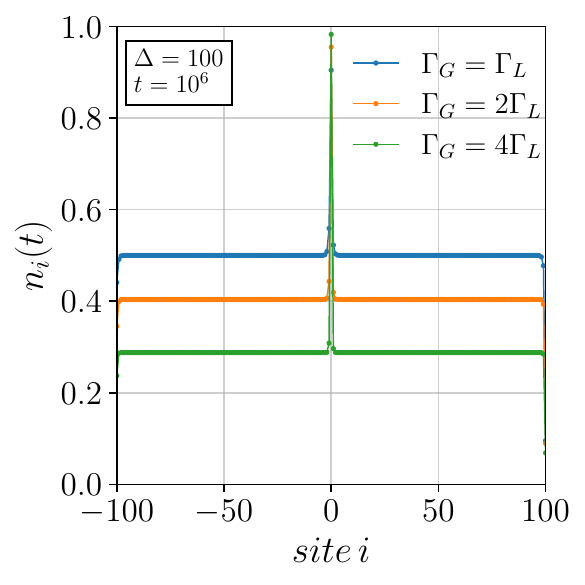}}
\caption{Local particle density profiles $n_i(t)$ [Eq.~\eqref{general_source_sink_wod_mat}] as a function of the lattice site $i$ at time $t=10^6$ for a system of size $L=201$, with a localized source at the origin and a localized sink separated by a distance $\Delta$. The initial state is a partially filled Fermi sea with filling $N(0)/L=0.25$ [Eq.~\eqref{ini_partial}]. Panels correspond to (a) $\Delta=20$, (b) $\Delta=50$, and (c) $\Delta=100$. In panels (a) and (b), secondary density peaks emerge at integer multiples of $\Delta$, demonstrating the robustness of these coherent interference features. In contrast, no secondary peaks appear for $\Delta=100$ [panel (c)] due to finite-size constraints under periodic boundary conditions ($L=201$), which truncate spatial interference before secondary peaks can form. The loss rate $\Gamma_L = J = 1$ for all curves.}
\label{fig:diff_delta_wod1}
\end{figure*}

To systematically investigate the origin and robustness of the secondary density peaks, we varied the source--sink separation $\Delta$ alongside the source and sink strengths. Figure~\ref{fig:diff_delta_wod1} presents the steady-state density profiles for the Fermi sea initial condition [Eq.~\eqref{ini_partial}] across different values of $\Delta$ and source strengths $\Gamma_G$. 

Notably, secondary peaks are absent for $\Delta=100$, as shown in Fig.~\ref{fig:diff_delta_wod1}(c). This is a finite-size effect resulting from periodic boundary conditions on a lattice of size $L=201$. Because the separation $\Delta$ is nearly half the total system length, spatial interference lacks the physical extent needed to develop secondary peaks before encountering the boundary. Corresponding results obtained by varying the sink strength $\Gamma_L$ are provided in Fig.~\ref{fig:diff_delta_wod2} of the Supplemental Material. 

Across parameter space, the density profiles exhibit consistent qualitative structure: secondary peaks consistently occur at integer multiples of $\Delta$, while their relative amplitudes are governed by local dissipation rates. The persistence of these peaks confirms that they are robust manifestations of quantum interference during coherent evolution rather than artifacts of specific parameter selections.

\subsection{Localized sink and source with dephasing}
In the presence of local dephasing, the dynamics is governed by the Lindblad master equation
\begin{align}
\partial_t \rho = -i[H,\rho]
+ \mathcal{L}_{\mathrm{sink}} \rho
+ \mathcal{L}_{\mathrm{source}} \rho 
+ \mathcal{L}_{\mathrm{deph}} \rho,
\label{lind_sink_source_wd}
\end{align}   
where $\mathcal{L}_{\mathrm{source}} \rho$, $\mathcal{L}_{\mathrm{sink}} \rho$, and $\mathcal{L}_{\mathrm{deph}} \rho$ are defined by Eqs.~\eqref{source_diss1}, \eqref{sink_diss1}, and \eqref{dephase_ope}, respectively.

Using Eq.~\eqref{lind_sink_source_wd}, the equation of motion for the two-point correlation function is
\begin{align}
 \frac{d}{dt}C_{i,j} &= -iJ\big(C_{i-1,j} + C_{i+1,j} - C_{i,j+1} - C_{i,j-1}\big)  \nonumber \\
&\quad - \Gamma_G(\delta_{i,0} + \delta_{j,0})\,C_{i,j}  -  \Gamma_L(\delta_{i,\Delta} + \delta_{j,\Delta})\,C_{i,j} \nonumber \\ 
&\quad + 2\gamma(\delta_{i,j} - 1)C_{i,j} + 2\Gamma_G \delta_{i,0} \delta_{j,0}.   \label{EOM_source_sink_wd1}
\end{align}
Following the vectorization procedure used in Section~\ref{source}, Eq.~\eqref{EOM_source_sink_wd1} can be written in compact form as
\begin{align}
    \dot{C} = MC + P,
    \label{EOM_source_sink_wd2}
\end{align}
where $C(t)$ is the vectorized correlation matrix, $M$ is the $L^2 \times L^2$ matrix containing all homogeneous terms, and $P$ is the $L^2$-dimensional source vector containing the single inhomogeneous term $2\Gamma_G \delta_{i,0}\delta_{j,0}$.

The formal solution to Eq.~\eqref{EOM_source_sink_wd2} is given by
\begin{align}
    C(t)
    =
    e^{Mt}C(0)
    +
    \int_0^t d\tau \,
    e^{M(t-\tau)}P.
    \label{EOM_source_sink_wd3}
\end{align}

Diagonalizing $M=V\Lambda_M V^{-1}$, the local density profile is
\begin{align}
n_i(t)
&=
\sum_{\alpha,\beta=1}^{L^2}
e^{\lambda_\alpha^M t}
V_{r_i,\alpha}
V^{-1}_{\alpha,\beta}
C_\beta(0)
\nonumber\\
&\quad +
\sum_{\alpha,\beta=1}^{L^2}
\left[
\frac{e^{\lambda_\alpha^M t}-1}
{\lambda_\alpha^M}
\right]
V_{r_i,\alpha}
V^{-1}_{\alpha,\beta}
P_\beta,
\label{density_source_sink_wd}
\end{align}
where $r_i = i + (i-1)L$ maps the diagonal elements of the $L \times L$ correlation matrix onto the vectorized index.

\begin{figure*}[t!]
\centering
\subfloat[]{%
\includegraphics[width=0.32\textwidth]{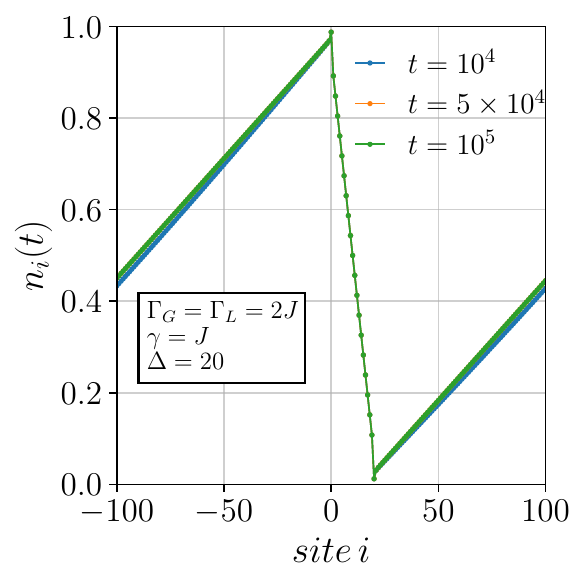}}
\hfill
\subfloat[]{%
\includegraphics[width=0.32\textwidth]{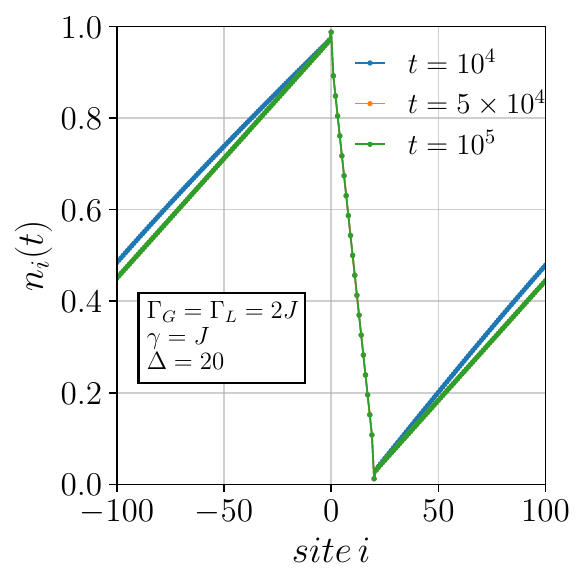}}
\hfill
\subfloat[]{%
\includegraphics[width=0.32\textwidth]{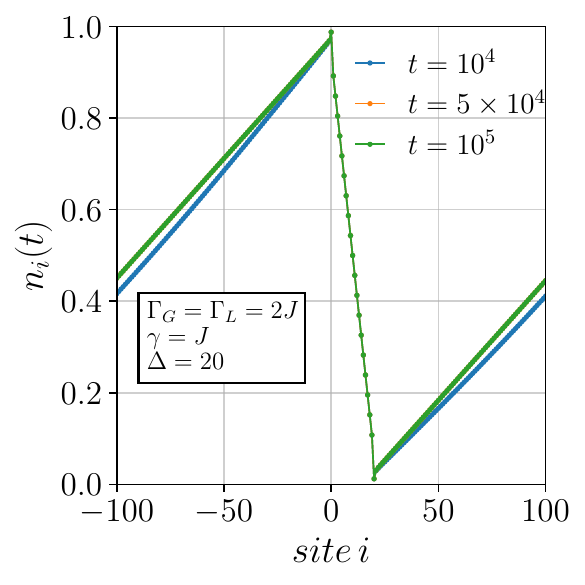}}
\caption{Local particle density profiles $n_i(t)$ [Eq.~\eqref{density_source_sink_wd}] as a function of the lattice site $i$ at different times for a system of size $L=201$, with a localized source at the origin, a localized sink located at a distance $\Delta=20$, and local dephasing of strength $\gamma$ acting on every lattice site. The three panels correspond to different initial conditions: (a) a partially filled Fermi sea with filling $N(0)/L=0.25$ [Eq.~\eqref{ini_partial}], (b) an initially fully occupied lattice [Eq.~\eqref{ini_full}], and (c) an initially empty lattice [Eq.~\eqref{ini_empty}]. In contrast to coherent dynamics, where quantum interference generates secondary density peaks, dephasing suppresses phase coherence and eliminates spatial interference structures. Consequently, irrespective of the initial condition, the system evolves to a unique nonequilibrium steady state characterized by a linear density profile governed by the diffusion equation. For all curves $J = 1$}
\label{fig:delta20_wd_difft}
\end{figure*}

\begin{figure*}[t!]
\centering
\subfloat[]{%
\includegraphics[width=0.32\textwidth]{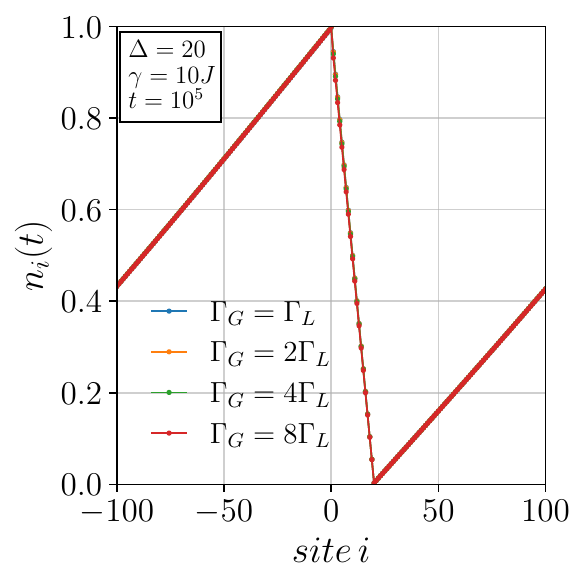}}
\hfill
\subfloat[]{%
\includegraphics[width=0.32\textwidth]{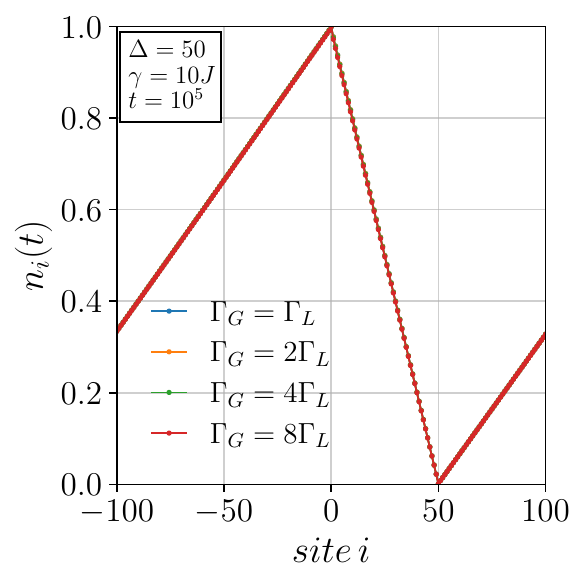}}
\hfill
\subfloat[]{%
\includegraphics[width=0.32\textwidth]{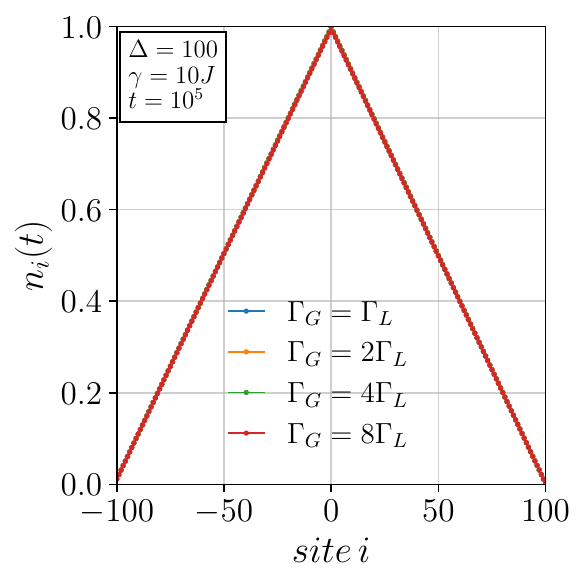}}
\caption{Local particle density profiles $n_i(t)$ [Eq.~\eqref{density_source_sink_wd}] as a function of the lattice site $i$ at time $t=10^5$ for a system of size $L=201$, with a localized source at the origin, a localized sink separated by a distance $\Delta$, and local dephasing of strength $\gamma$ acting on every lattice site. Panels correspond to different source--sink separations: (a) $\Delta=20$, (b) $\Delta=50$, and (c) $\Delta=100$. In contrast to the coherent case where secondary density peaks emerge at integer multiples of $\Delta$, dephasing suppresses quantum coherence across all values of $\Delta$. As a result, the density profiles evolve smoothly toward the classical diffusive regime. The initial state is a partially filled Fermi sea with filling $N(0)/L=0.25$ [Eq.~\eqref{ini_partial}]. Parameters are set to $\Gamma_L=J = 1$ for all curves.}
\label{fig:dephase_diffG}
\end{figure*}

The density profiles for $\Delta \neq 0$ across the three initial conditions are shown in Fig.~\ref{fig:delta20_wd_difft}. Unlike the coherent case, the secondary density peaks disappear entirely when local dephasing is introduced. Figure~\ref{fig:dephase_diffG} illustrates the steady-state density profiles across varying separations $\Delta$ and source strengths $\Gamma_G$. Local dephasing suppresses quantum coherence, eliminating spatial interference patterns and driving the system into the classical diffusive regime, where the steady-state density profile is governed by the macroscopic diffusion equation.
 
%%%%%%%%%%%%%%%%%%%%%%%%%%%%%%%%%%%%%%%%%%%%%%%%%%%%%%%%%%%%%%%%%%%%%%%%%%%%%%%%%%%%%%%%%
\section{Conclusion and Discussion}\label{conclusions}

We investigated the nonequilibrium transport of a one-dimensional spinless fermionic lattice coupled to localized particle sources and sinks in the presence of local dephasing. Within the Lindblad master equation framework, we computed the exact time evolution of two-point spatial correlation functions and local particle density profiles. Our results demonstrate that local dephasing serves as a control parameter driving a smooth crossover from ballistic quantum transport to purely classical diffusion.

In the single-defect setup (source or sink only), the particle growth or loss dynamics exhibits three distinct dynamical stages governed by ballistic propagation and boundary-induced interference. A central focus of this work is the spatial structure of the steady-state density profile generated under non-Markovian driving. In the purely unitary limit (zero dephasing), a localized sink generates persistent Friedel-type oscillations in the local density, reflecting the non-local response of the Fermi sea to localized loss. As local dephasing strength increases, phase coherence is systematically suppressed, washing out Friedel oscillations and collapsing the intermediate dynamical stage to yield a two-regime incoherent transport process.

When both a source and a sink are present at separation $\Delta$, the steady state exhibits features beyond simple boundary modulations. Under coherent evolution, secondary density maxima and minima emerge at integer multiples of $\Delta$ across the lattice. These secondary peaks are direct signatures of phase-coherent quantum propagation and multipath interference under unitary hopping. Introducing local dephasing suppresses these interference-induced spatial structures, restoring a linear density profile governed by classical bulk diffusion (see Figs.~\ref{fig:delta20_wd_difft} and \ref{fig:dephase_diffG}).

Looking forward, an exciting direction is to investigate higher cumulants and full counting statistics (FCS) of particle transport in this setup to further quantify quantum-to-classical noise cross-overs under localized dissipation.

\section{Acknowledgments}
We acknowledge useful discussions with Bijay Agarwalla, Manas Kulkarni, and Anupam Kundu. E.B.\ thanks Rajiv G Pereira for insightful discussions, Subhodeep Dey for valuable comments, and Arpan Chatterjee, Lakshmi N Govind, Arnab Mandal, and Sayan Sircar for helpful interactions. S.S.\ thanks Katha Ganguly for useful discussions. This project was funded by intramural funds at TIFR Hyderabad from the Department of Atomic Energy (DAE), Government of India, under Project Identification No.~RTI4007.

\appendix 

\section{Analytical results in the strong dephasing limit}
\label{A1}
In the strong dephasing limit, the continuum evolution equation for the particle density $n(x,t)$ with a sink located at $x_0 = 0$ is given by
\begin{align}
    \frac{\partial}{\partial t}n(x,t) = \omega_1 \frac{\partial^2}{\partial x^2} n(x,t) - 2\omega_1 \delta(x) n(x,t), \label{a_continuum_limit}
\end{align}
subject to the initial condition $n(x,0) = n_0(x)$ and vanishing boundary conditions $n(x \to \pm\infty, t) = 0$.

Taking the Laplace transform with respect to time,
\begin{align}
    \tilde{n}(x,s) = \int_{0}^{\infty} dt \, e^{-st} \, n(x,t), \label{a_def_laplace}
\end{align}
Eq.~\eqref{a_continuum_limit} becomes
\begin{align}
    s \,\tilde{n}(x,s) - n_0(x) = \omega_1 \frac{\partial^2}{\partial x^2} \tilde{n}(x,s) - 2\omega_1 \delta(x) \tilde{n}(x,s). \label{a_laplace_form}
\end{align}
Applying a spatial Fourier transform,
\begin{align}
    n(k,s) = \int_{-\infty}^{\infty} dx \, e^{-ikx} \, \tilde{n}(x,s), \label{a_def_fourier}
\end{align}
recasts Eq.~\eqref{a_laplace_form} into
\begin{align}
    s \, n(k,s) - n_0(k) = -\omega_1 k^2 n(k,s) - 2 \omega_1 \tilde{n}(0,s), \label{fourier_form}
\end{align}
yielding
\begin{align}
    n(k,s) = \big[n_0(k) - 2\omega_1\, \tilde{n}(0,s)\big]\psi(s,k) \label{a_laplace_fourier}
\end{align}
with

\begin{align*}
  \psi(s,k) = \frac{1}{s + \omega_1 k^2}  
\end{align*}

From the inverse relation of Eq.~\eqref{a_def_fourier},
\begin{align}
    \tilde{n}(0,s) = \frac{1}{2\pi} \int_{-\infty}^{\infty} dk \, n(k,s), \label{a_fourier_zeroth}
\end{align}
which, in conjunction with Eq.~\eqref{a_laplace_fourier}, determines the local density at the origin:
\begin{align}
    \tilde{n}(0,s) = \frac{k_F}{\pi\big(s + \sqrt{s\omega_1}\big)}. \label{a_zeroth_laplace}
\end{align}

Here, $n_0(k)$ is the Fourier transform of the initial density $n_0(x) = \langle a^\dagger(x) a(x) \rangle$. Expressing the field operators in momentum space as
\begin{align}
    a(x) = \frac{1}{\sqrt{2\pi}} \int_{-\infty}^{\infty} dk \, e^{ikx} \, \tilde{a}_k, \label{a_density_fourier}
\end{align}
and utilizing the zero-temperature expectation value $\langle \tilde{a}^\dagger_k(0) \tilde{a}_{k'}(0) \rangle = 2\pi \delta(k - k') \Theta(k_F^2 - k^2)$, the initial momentum-space density profile evaluates to
\begin{align}
    n_0(k) = 2k_F \delta(k).
\end{align}

Substituting $n_0(k)$ and Eq.~\eqref{a_zeroth_laplace} back into Eq.~\eqref{a_laplace_fourier}, we obtain
\begin{align}
    n(k,s) = \frac{n_0(k)}{s + \omega_1 k^2} - \left[\frac{k_F}{\pi (s + \sqrt{\omega_1 s})}\right] \left(\frac{2\omega_1}{s + \omega_1 k^2}\right). \label{a_laplace_fourier2}
\end{align}

Applying the inverse spatial Fourier transform,
\begin{align}
    \tilde{n}(x,s) = \frac{1}{2\pi} \int_{-\infty}^{\infty} dk \, e^{ikx} \, n(k,s), \label{a_inverse_FT}
\end{align}
gives
\begin{align}
\tilde{n}(x,s)
&= \frac{1}{2\pi}
\int_{-\infty}^{\infty} dk \,
e^{ikx}
\frac{2k_F\delta(k)}{s+\omega_1k^2}
\nonumber\\
&\quad
-\frac{\omega_1 k_F}
{\pi^2\left(s+\sqrt{\omega_1 s}\right)}
\int_{-\infty}^{\infty} dk \,
\frac{e^{ikx}}{s+\omega_1k^2}
\nonumber\\
&= \frac{k_F}{\pi s}
-\frac{k_F\sqrt{\omega_1}}
{\pi^2\sqrt{s}\left(s+\sqrt{\omega_1 s}\right)}
e^{-|x|\sqrt{s/\omega_1}}.
\label{a_eq40}
\end{align}

Finally, performing the inverse Laplace transform of Eq.~\eqref{a_eq40} yields
\begin{align}
    n(x,t) &= \frac{k_F}{\pi} - \frac{k_F}{\pi}\,\operatorname{erfc}\left(\frac{|x|}{2\sqrt{\omega_1 t}}\right) \nonumber \\
    &+ \frac{k_F}{\pi}\,\exp\left(|x| + \omega_1 t\right)\,\operatorname{erfc}\left(\frac{2\omega_1 t + |x|}{2\sqrt{\omega_1 t}}\right). \label{a_41}
\end{align}
Using the relation $\operatorname{erf}(z) = 1 - \operatorname{erfc}(z)$, Eq.~\eqref{a_41} simplifies to the exact analytical density solution:
\begin{align}
    n(x,t) = \frac{k_F}{\pi}\,\operatorname{erf}\left(\frac{|x|}{2\sqrt{\omega_1 t}}\right) + \frac{k_F}{\pi}\,\exp\left(|x| + \omega_1 t\right)\,\operatorname{erfc}\left(\frac{2\omega_1 t + |x|}{2\sqrt{\omega_1 t}}\right). \label{a_42}
\end{align}

%%%%%%%%%%%%%%%%%%%%%%%%%%%%%%%%%%%%%%%%
\bibliographystyle{apsrev4-2}
\bibliography{bibfile.bib}
%%%%%%%%%%%%%%%%%%%%%%%%%%%%%%%%%%%%%%%%

%%%%%%%%%%%%%%%%%%%%%%%%%%%%%%%%%%%%%%%%%%%%%%%%%%%%%%%%%
%%%%%%%%%%%%%%%%%%%%%%%%%%%%%%%%%%%%%%%%%%%%%%%%%%%%%%%%%

\clearpage
\onecolumngrid
\setcounter{equation}{0}

\section*{Supplemental Material}

\subsection{Friedel Oscillations}
The density matrix $\rho$ of the system evolving in the presence of a localized lossy site obeys the Lindblad master equation
\begin{align}
\partial_t \rho = -i[H,\rho] + \Gamma_L\big(2 a_0 \rho a^{\dagger}_0 - \{a^{\dagger}_0 a_0,\rho\}\big).
\label{sup_sink}
\end{align}
Equation~\eqref{sup_sink} can be recast as
\begin{align}
\partial_t \rho = -i(H_{\mathrm{eff}}\rho - \rho H^{\dagger}_{\mathrm{eff}}) + 2 \Gamma_L a^{\dagger}_0 \rho a_0,
\end{align}
where
\begin{align}
H_{\mathrm{eff}} = H - i\Gamma_L a^{\dagger}_0 a_0
\end{align}
is an effective non-Hermitian Hamiltonian. Here, $H_{\mathrm{eff}}$ represents a system subject to a localized imaginary potential at site $i=0$. We compute the retarded Green's function for the non-interacting Hamiltonian with an imaginary potential at the origin.

The retarded Green's function for the free tight-binding Hamiltonian in momentum space is
\begin{align}
G^0(k,\omega) = \frac{1}{\omega + i 0^+ - \varepsilon_k}, \quad \varepsilon_k = -2J\cos k. \label{free_G}
\end{align}
Taking the spatial Fourier transform of Eq.~\eqref{free_G}, the real-space unperturbed Green's function takes the form
\begin{align}
G_{i-j}^0(\omega) = \frac{\left(-\frac{\omega}{2J} + i\sqrt{1-\frac{\omega^2}{4J^2}}\right)^{|i-j|}}{2iJ\sqrt{1-\frac{\omega^2}{4J^2}}}.
\end{align}

Following the Dyson equation, the full retarded Green's function satisfies
\begin{align}
G = G^0 + G^0 V G,
\label{dyson}
\end{align}
where the non-Hermitian potential is
\begin{align}
V = - i\Gamma_L a^{\dagger}_0 a_0.
\label{im_V}
\end{align}
Using Eq.~\eqref{dyson} with the localized imaginary potential of Eq.~\eqref{im_V}, the retarded Green's function in momentum space evaluates to
\begin{align}
G(k,k',\omega) = G^0(k,\omega)\delta_{k,k'} - \frac{i \Gamma_L G^0(k,\omega) \, G^0(k',\omega)}{1+i\Gamma_L G_{j=0}^0(\omega)}.
\label{new_G}
\end{align}

The two-point, two-time correlation function in real space is given by
\begin{align}
C_{i,j}(t,t') = \sum_{n,n'} G_{i,n}^*(t-t_0) \, G_{j,n'}(t'-t_0) \, C_{n,n'}(t_0,t_0), \label{green_corr}
\end{align}
where the initial correlation matrix $C_{n,n'}(t_0,t_0)$ for a translationally invariant system is
\begin{align}
C_{n,n'}(t_0,t_0) = \frac{1}{L} \sum_k e^{-ik(n-n')} n_{0,k}. \label{sup_ini_corr}
\end{align}
Substituting Eq.~\eqref{sup_ini_corr} into Eq.~\eqref{green_corr}, the correlation function simplifies to
\begin{align}
C_{i,j}(t,t') = \frac{1}{L} \sum_k G_i^*(k, t-t_0) \, G_j(k, t'-t_0) \, n_{0,k}.
\label{G_one}
\end{align}

In mixed real- and momentum-space representations, the retarded Green's function is
\begin{align}
G_{i,j}(\omega) = \frac{1}{L} \sum_{k,k'} e^{iki} e^{-ik'j} G(k,k',\omega),
\label{real_new_G}
\end{align}
and
\begin{align}
G_j(k,\omega) = \sum_i e^{-iki} \, G_{i,j}(\omega) = \sum_q e^{-iqj} \, G(k,q, \omega).
\label{mixed_G2}
\end{align}
Evaluating Eq.~\eqref{mixed_G2} using Eq.~\eqref{new_G} yields \cite{Ultracold}
\begin{align}
G_j(k,\omega) = G^0(k,\omega) \left[e^{-ikj} + r(\omega) \, e^{i g(\omega) |j|}\right],
\label{mixed_G3}
\end{align}
where the effective reflection amplitude $r(\omega)$ and the effective momentum $g(\omega)$ are defined as
\begin{align}
r(\omega) = \frac{\Gamma_L}{\Gamma_L + v(\omega)}, \quad v(\omega) = J \sqrt{4 - \frac{\omega^2}{J^2}},
\end{align}
and
\begin{align}
g(\omega) = -i\ln\left(-\frac{\omega}{2J} + i \sqrt{1 - \frac{\omega^2}{4J^2}}\right).
\end{align}

Substituting Eq.~\eqref{mixed_G3} into Eq.~\eqref{G_one}, the equal-time density matrix elements $C_{i,j}(t,t)$ become
\begin{align}
C_{i,j}(t,t) = \frac{1}{L} \sum_k f^*_i(\varepsilon_k,k) \, f_j(\varepsilon_k, k) \, n_{0,k},
\end{align}
with $f_j(\varepsilon_k,k) = e^{-ikj} + r(\varepsilon_k) e^{i g(\varepsilon_k) |j|}$.

The local particle density is therefore
\begin{align}
C_{j,j}(t) = \frac{1}{L} \sum_k |f_j(\varepsilon_k,k)|^2 \, n_{0,k}.
\label{sup_corr1}
\end{align}
In the thermodynamic limit ($L \to \infty$) at zero temperature, where $n_{0,k} = \Theta(k_F^2 - k^2)$, Eq.~\eqref{sup_corr1} converts into the integral
\begin{align}
C_{j,j}(t) = \frac{1}{2\pi} \int_{-k_F}^{k_F} dk \, |f_j(\varepsilon_k,k)|^2.
\end{align}
Splitting the integral over positive and negative momentum sectors yields
\begin{align}
C_{j,j}(t) &= \frac{1}{\pi} \int_{0}^{k_F} dk \, \Big[1 + r_k + r^2_k + r_k \cos(2k|j|)\Big] \nonumber \\
&= \frac{1}{\pi} \int_{0}^{k_F} dk \, (1 + r_k + r^2_k) + \frac{1}{\pi} \int_{0}^{k_F} dk \, r_k \cos(2k|j|) \nonumber \\
&\equiv n_{\mathrm{bg}} + \delta n,
\end{align}
where the spatially oscillating component $\delta n$ evaluates asymptotically to
\begin{align}
\delta n \approx \frac{1}{\pi} r_{k_F} \frac{\sin(2k_F |j|)}{2|j|}. \label{friedel}
\end{align}
Equation~\eqref{friedel} represents the steady-state Friedel oscillations induced by the localized sink, characterized by the characteristic wavevector $2k_F$.

\subsection{Anti-Quantum Zeno Effect (AQZE) to Quantum Zeno Effect (QZE) Crossover}

To analyze the effect of local dephasing $\gamma$ on particle loss dynamics, we evaluate the net particle loss rate $|dN/dt|$ in the second dynamical regime as a function of $\gamma/J$. 

At low initial filling fractions $n_0 = N(0)/L$, the net loss rate initially increases with $\gamma$, reaching a maximum characteristic of the Anti-Quantum Zeno Effect (AQZE). Beyond this peak, further increasing $\gamma$ suppresses the loss rate, signaling the onset of the standard Quantum Zeno Effect (QZE). 

\begin{figure*}[htbp]
\centering
\subfloat[]{%
\includegraphics[width=0.33\textwidth]{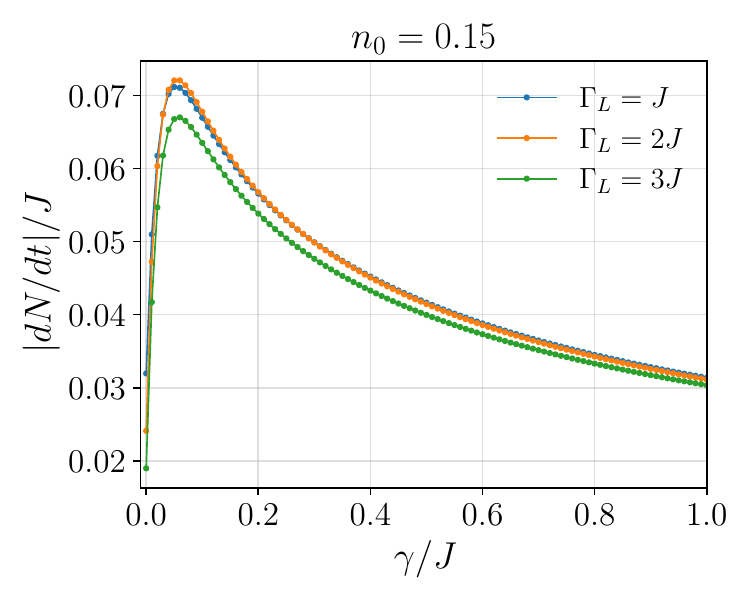}}
\hfill
\subfloat[]{%
\includegraphics[width=0.33\textwidth]{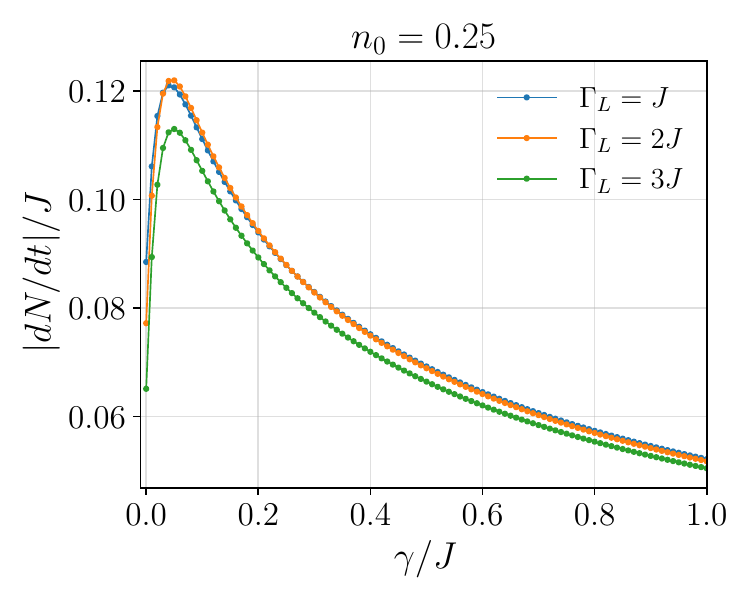}}
\hfill
\subfloat[]{%
\includegraphics[width=0.33\textwidth]{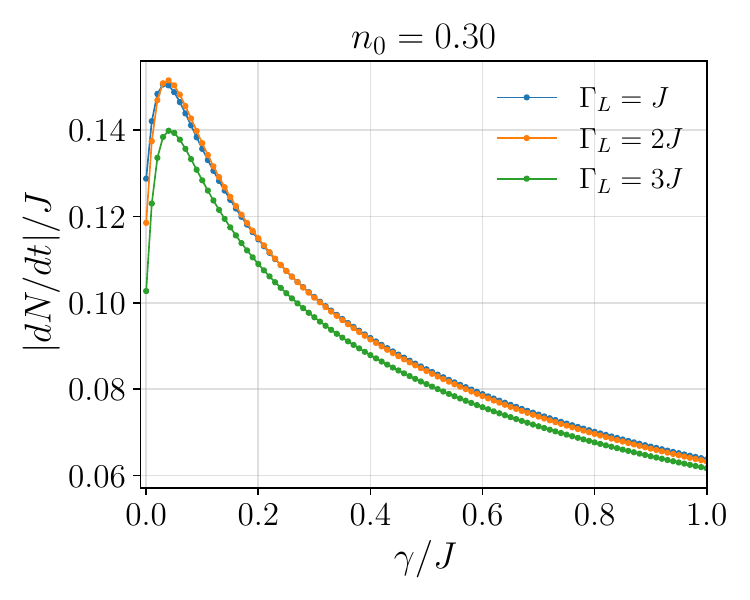}}
\caption{Particle loss rate $|dN/dt|/J$ in the second dynamical regime as a function of the dephasing rate $\gamma/J$ for various sink strengths $\Gamma_L$ and low initial filling fractions: (a) $n_0 = 0.15$, (b) $n_0 = 0.25$, and (c) $n_0 = 0.30$. Increasing the density progressively diminishes the AQZE peak.}
\label{fig:QZE_sink_1}
\end{figure*}

\begin{figure*}[htbp]
\centering
\subfloat[]{%
\includegraphics[width=0.45\textwidth]{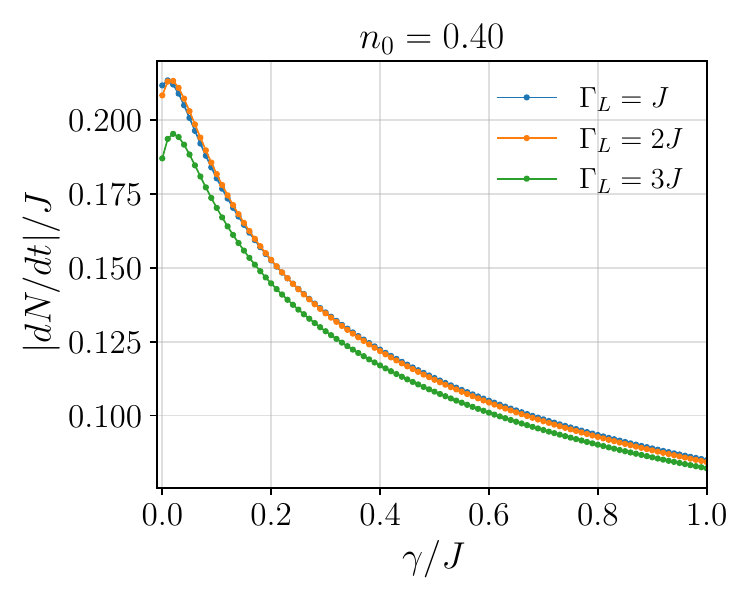}}
\hfill
\subfloat[]{%
\includegraphics[width=0.45\textwidth]{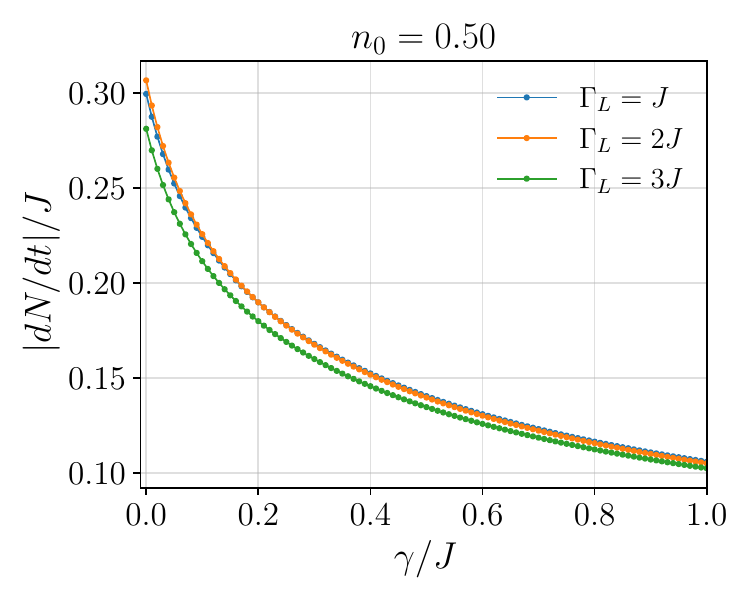}}
\caption{Particle loss rate $|dN/dt|/J$ in the second dynamical regime as a function of the dephasing rate $\gamma/J$ for higher filling fractions: (a) $n_0 = 0.40$ and (b) $n_0 = 0.50$ (half-filling). The AQZE peak is substantially suppressed at $n_0 = 0.40$ (a) and completely disappears at half-filling $n_0 = 0.50$ (b), leaving purely QZE behavior.}
\label{fig:QZE_sink_2}
\end{figure*}

However, as the initial filling $n_0$ increases, the AQZE regime is progressively suppressed due to spatial Fermi surface constraints and quantum interference among filled states. At high densities, the AQZE maximum vanishes completely, leaving only the monotonic QZE decay. These behaviors are illustrated in Figs.~\ref{fig:QZE_sink_1} and \ref{fig:QZE_sink_2}.

\newpage
\subsection{Relation between Measurements and the Lindbladian}

Measurements are described by a set of measurement operators $\{M_m\}$. If the state before measurement is $|\psi\rangle$, the probability of obtaining outcome $m$ is
\begin{align}
    p(m) = \langle \psi|M^{\dagger}_m M_m|\psi\rangle.
\end{align}
The state immediately after measurement becomes
\begin{align}
    |\psi^m\rangle = \frac{M_m|\psi\rangle}{\sqrt{\langle \psi|M^{\dagger}_m M_m|\psi\rangle}}.
\end{align}
These measurement operators satisfy the completeness relation $\sum_m M^{\dagger}_m M_m = \mathbb{I}$, which guarantees that total probability is conserved:
\begin{align}
    \sum_m p(m) = \sum_m \langle\psi|M^{\dagger}_mM_m|\psi\rangle = 1.
\end{align}

For a mixed state described by a density matrix $\rho = \sum_i p_i |\psi_i\rangle \langle\psi_i|$ (where $\sum_i p_i = 1$), the conditional probability of obtaining outcome $m$ given state $|\psi_i\rangle$ is
\begin{align}
    p(m|i) = \langle \psi_i|M^{\dagger}_m M_m|\psi_i \rangle.
\end{align}
By the law of total probability, the unconditional probability of measuring $m$ is
\begin{align}
    p(m) &= \sum_i p(m|i) p_i = \sum_i p_i \langle \psi_i|M^{\dagger}_m M_m|\psi_i \rangle \nonumber \\
    &= \sum_i p_i \text{Tr}\left(M^{\dagger}_m M_m |\psi_i\rangle \langle\psi_i|\right) \nonumber \\
    &= \text{Tr}(M_m \rho M^{\dagger}_m).
\end{align}

If outcome $m$ is observed, the post-measurement state associated with component $|\psi_i\rangle$ is
\begin{align}
    |\psi^m_i\rangle = \frac{M_m|\psi_i\rangle}{\sqrt{\langle \psi_i|M^{\dagger}_m M_m|\psi_i\rangle}}.
\end{align}
Using Bayes' rule $p(i|m) = \frac{p(m|i)p_i}{p(m)}$, the conditional density matrix given outcome $m$ is
\begin{align}
    \rho_m &= \sum_i p(i|m) |\psi^m_i\rangle \langle\psi^m_i| \nonumber \\
    &= \sum_i \frac{p(m|i) p_i}{p(m)} \frac{M_m|\psi_i\rangle \langle\psi_i|M^{\dagger}_m}{\langle \psi_i|M^{\dagger}_m M_m|\psi_i\rangle} \nonumber \\
    &= \frac{M_m \rho M^{\dagger}_m}{\text{Tr}(M_m \rho M^{\dagger}_m)}.
\end{align}
This represents the updated state conditioned on outcome $m$. Summing over all unobserved outcomes gives the updated unconditional density matrix:
\begin{align}
    \rho(t+dt) = \sum_m p(m) \rho_m = \sum_m M_m \rho M^{\dagger}_m. \label{measurement}
\end{align}

Now consider a system evolving according to the Quantum Master Equation in Lindblad form:
\begin{align}
    \frac{d\rho}{dt} = \mathcal{L}\rho = -i[H,\rho] + \sum_{k=1}^r \mathcal{D}[L_k]\rho, \quad \text{where } \mathcal{D}[L]\rho = L \rho L^{\dagger} - \frac{1}{2}\{L^{\dagger}L,\rho\}, \label{lindblad}
\end{align}
where $\{L_k\}$ are the jump operators for each decay/dephasing channel $k$. The formal solution to Eq.~\eqref{lindblad} is
\begin{align}
    \rho(t) = e^{\mathcal{L}t}\rho(0).
\end{align}

For an infinitesimal time step $dt$, we expand $e^{\mathcal{L}dt}\rho = \rho + dt\,\mathcal{L}\rho + \mathcal{O}(dt^2)$:
\begin{align}
    e^{\mathcal{L}dt}\rho &= \rho - i dt [H,\rho] + dt \sum_k \left( L_k \rho L_k^\dagger - \frac{1}{2}\{L_k^\dagger L_k, \rho\} \right) \nonumber \\
    &= \rho - i dt (H_{\text{eff}}\rho - \rho H^{\dagger}_{\text{eff}}) + dt \sum_{k=1}^r L_k \rho L^{\dagger}_k + \mathcal{O}(dt^2), \label{lind_measure}
\end{align}
where $H_{\text{eff}} = H - \frac{i}{2}\sum_{k} L^{\dagger}_k L_k$ is an effective non-Hermitian Hamiltonian. 

Eq.~\eqref{lind_measure} can be cast in the form of a quantum channel/Kraus map $\rho(t+dt) = \sum_k M_k \rho M_k^\dagger$ with Kraus operators:
\begin{align}
    M_0 &= \mathbb{I} - i H_{\text{eff}} dt, \\
    M_k &= \sqrt{dt}\, L_k, \quad (k = 1, 2, \dots, r).
\end{align}
These operators satisfy completeness up to first order in $dt$:
\begin{align}
    M^{\dagger}_0 M_0 + \sum_{k=1}^r M^{\dagger}_k M_k = \mathbb{I} + \mathcal{O}(dt^2).
\end{align}

%%%%%%%%%%%%%%%%%%%%%%%%%%%%%%%%%%%%%%%%%%%%%%%%%%%%%%%%%
\begin{figure*}[h!]
\centering
\includegraphics[width=0.45\textwidth]{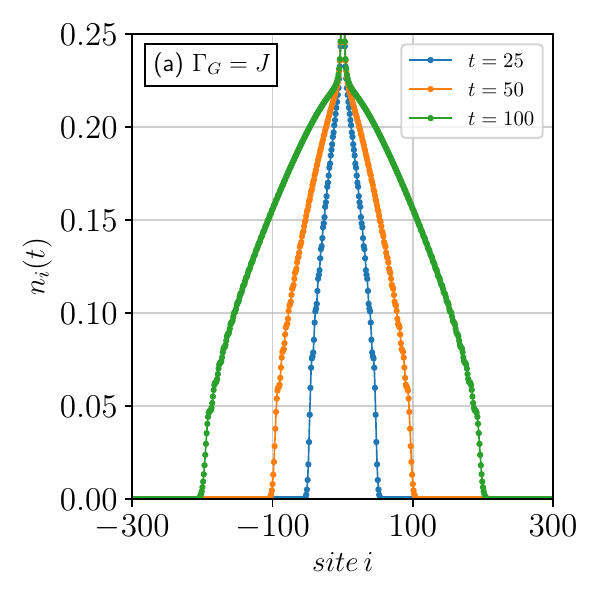}
\hfill
\includegraphics[width=0.45\textwidth]{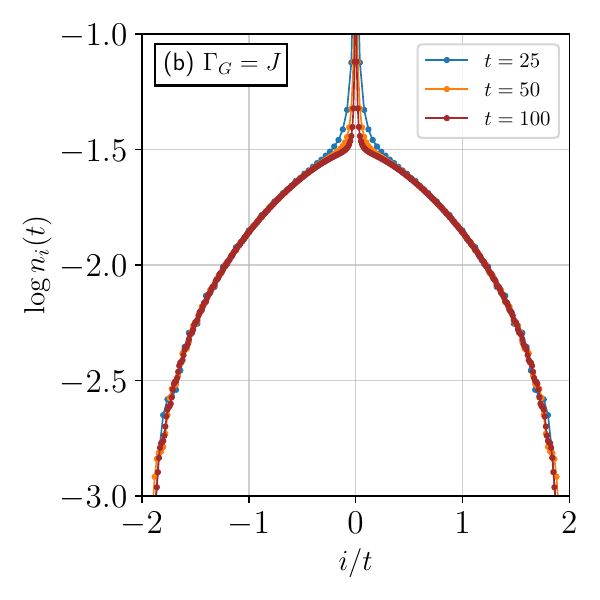}
\caption{
\textbf{(a)} Local density profile $n_i(t)$ (Eq.~\eqref{sup_source_wod}) vs.\ lattice site $i$ with a source of rate $\Gamma_G$ attached at the origin.
\textbf{(b)} Scaled density $\log n_i(t)$ vs.\ scaled coordinate $i/t$ at various times. The ballistic nature of the transport is confirmed by the collapse under $i/t$ scaling. Parameters for both plots: $L = 601$ and $J = 1$.
}
\label{fig:sup_source_wod}
\end{figure*}
%%%%%%%%%%%%%%%%%%%%%%%%%%%%%%%%%%%%%%%%%%%%%%%%%%%%%%%%%

\subsection{Localized Source Without Dephasing}

The formal solution for the two-point correlation function can be written as
\begin{align}
    C(t) = \int_{0}^{t} d\tau \, e^{G(t-\tau)} P e^{G^{\dagger}(t-\tau)},
\end{align}
where $G = iW + L$, and $W$, $L$, and $P$ are defined in the main text. The local density profile takes the form
\begin{align}
    n_i(t) = \sum_{\alpha,\beta} \left[ \frac{e^{(\lambda_{\alpha}^1 +\lambda_{\beta}^2)t }-1}{\lambda_{\alpha}^1 +\lambda_{\beta}^2} \right] V_{i\alpha} X_{\alpha \beta} U_{\beta i}^{-1}, \label{sup_source_wod}
\end{align}
where $X = V^{-1} P U$. Figure~\ref{fig:sup_source_wod}(a) shows the evolution of the density profile at different times, while Fig.~\ref{fig:sup_source_wod}(b) presents $\log n_i(t)$ as a function of $i/t$. The excellent collapse of the data demonstrates the ballistic scaling behavior.
%%%%%%%%%%%%%%%%%%%%%%%%%%%%%%%%%%%%%%%%%%%%%%%%%%%%%%%%%
\begin{figure*}[h!]
\centering
\includegraphics[width=0.45\textwidth]{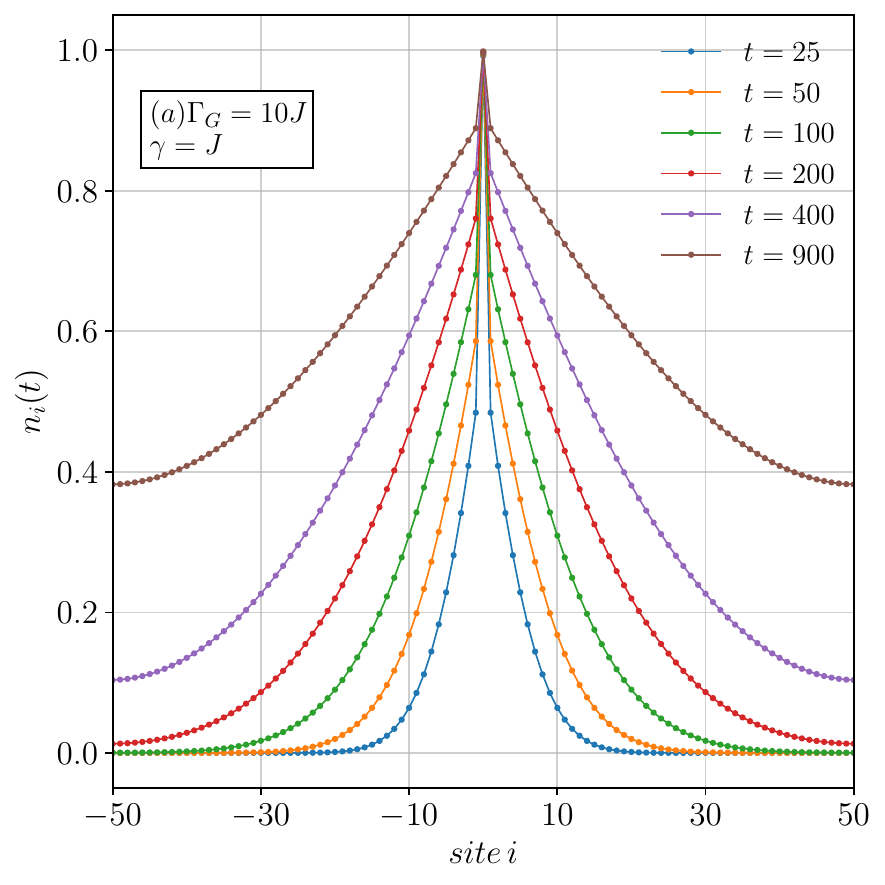}
\hfill
\includegraphics[width=0.45\textwidth]{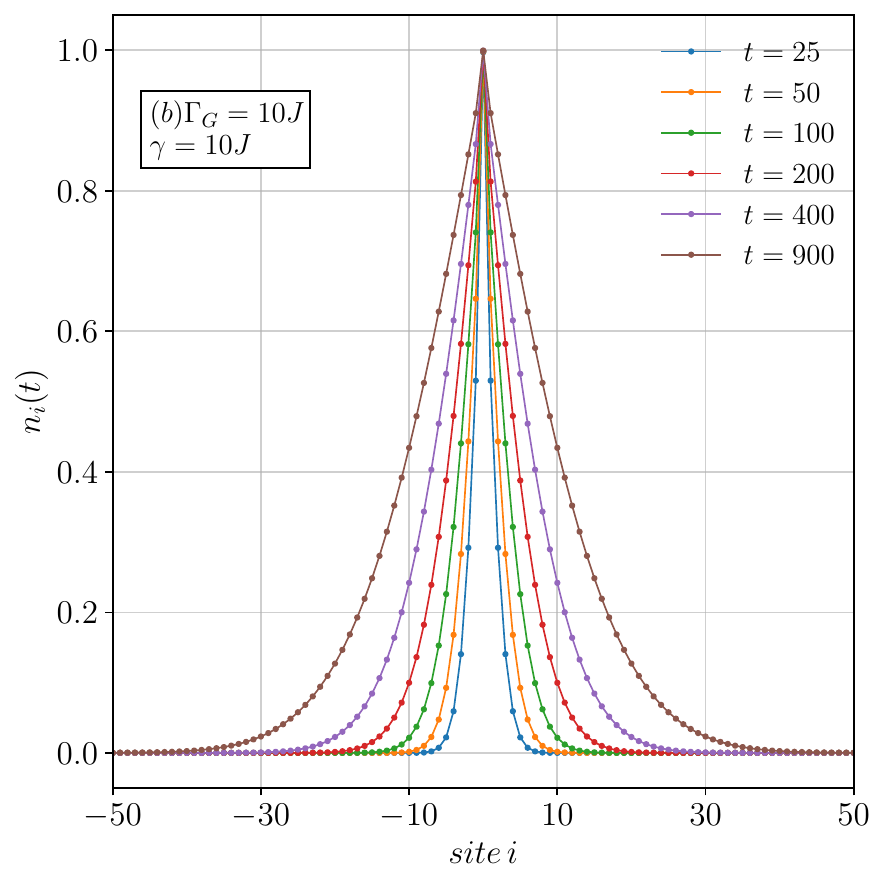}
\caption{
Local density profile $n_i(t)$ (Eq.~\eqref{sup_density_source_wd}) across lattice sites $i$, with source rate $\Gamma_G$ at the origin and dephasing rate $\gamma$ at all sites. 
\textbf{(a)} Strong source relative to dephasing ($\Gamma_G > \gamma$), leading to density buildup near the origin. 
\textbf{(b)} Equal source and dephasing strengths ($\Gamma_G = \gamma$), promoting spatial spreading. Parameters: $L = 101$, $J = 1$.
}
\label{fig:sup_source_wd_fig2}
\end{figure*}
%%%%%%%%%%%%%%%%%%%%%%%%%%%%%%%%%%%%%%%%%%%%%%%%%%%%%%%%%

%%%%%%%%%%%%%%%%%%%%%%%%%%%%%%%%%%%%%%%%%%%%%%%%%%%%%%%%%
\begin{figure*}[htbp]
\centering
\includegraphics[width=0.45\textwidth]{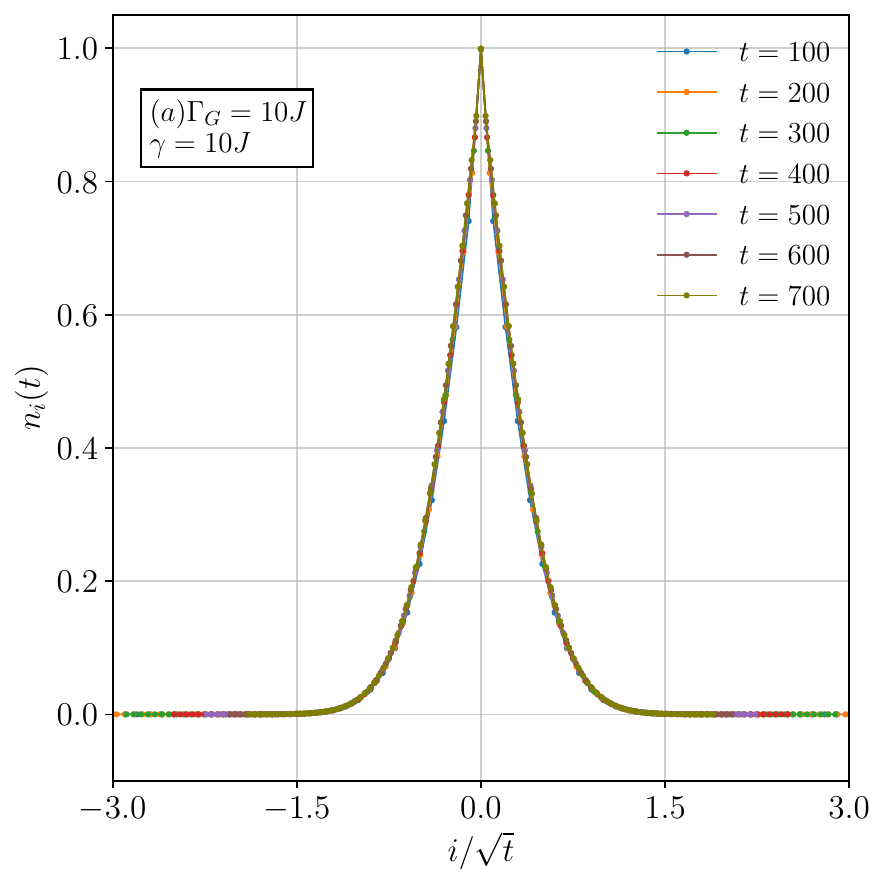}
\hfill
\includegraphics[width=0.45\textwidth]{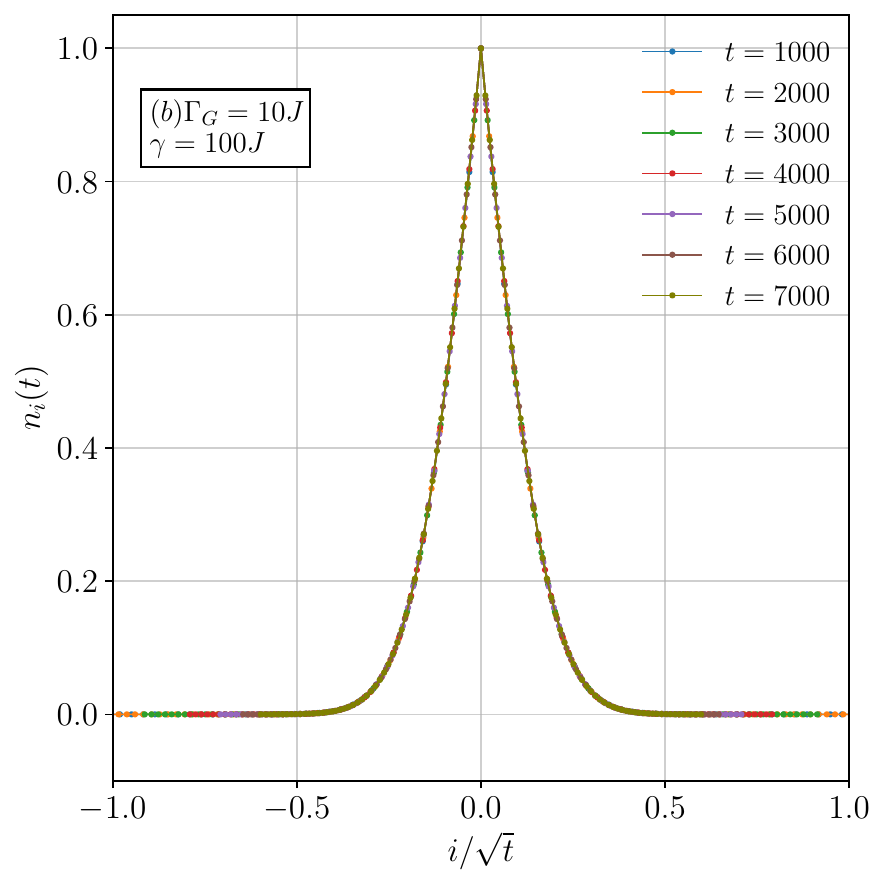}
\caption{
Density profile vs.\ scaled coordinate $i/\sqrt{t}$ at different times. 
\textbf{(a)} For $\Gamma_G \approx \gamma$, curve collapse indicates rapid crossover to diffusive transport. 
\textbf{(b)} For $\gamma \gg \Gamma_G$, the collapse occurs at much later times due to dephasing-induced slowdown. Parameters: $L = 101$, $J = 1$.
}
\label{fig:sup_scaled_source_wd}
\end{figure*}
%%%%%%%%%%%%%%%%%%%%%%%%%%%%%%%%%%%%%%%%%%%%%%%%%%%%%%%%%

\subsection{Localized Source With Dephasing}

In the presence of dephasing, the correlation function satisfies
\begin{align}
    C(t) = \int_{0}^{t} d\tau \, e^{M(t-\tau)} P, \label{sup_corr_source_wd}
\end{align}
where $M$ is the superoperator matrix. Vectorizing the correlation matrix, the $r$-th component $C_r(t)$ is given by
\begin{align}
    C_r(t) = \sum_{\alpha,\beta} \left[ \frac{e^{\lambda^M_{\alpha} t }-1}{\lambda^M_{\alpha}} \right] U_{r\alpha} U_{\alpha \beta}^{-1} P_{\beta}, \label{sup_rth_source_wd}
\end{align}
and the site density profile is extracted via
\begin{align}
    n_i(t) = C_{i+(i-1)L}(t). \label{sup_density_source_wd}
\end{align}

When the dephasing rate $\gamma$ is comparable to $\Gamma_G$, diffusive scaling ($i/\sqrt{t}$) sets in early. Conversely, when $\gamma \gg \Gamma_G$, Quantum Zeno-like suppression delays the onset of the diffusive regime; see Figs.~\ref{fig:sup_source_wd_fig2} and \ref{fig:sup_scaled_source_wd}.

\subsection{Localized Sink and Source Without Dephasing}

\begin{figure*}[htbp]
\centering
\includegraphics[width=0.32\textwidth]{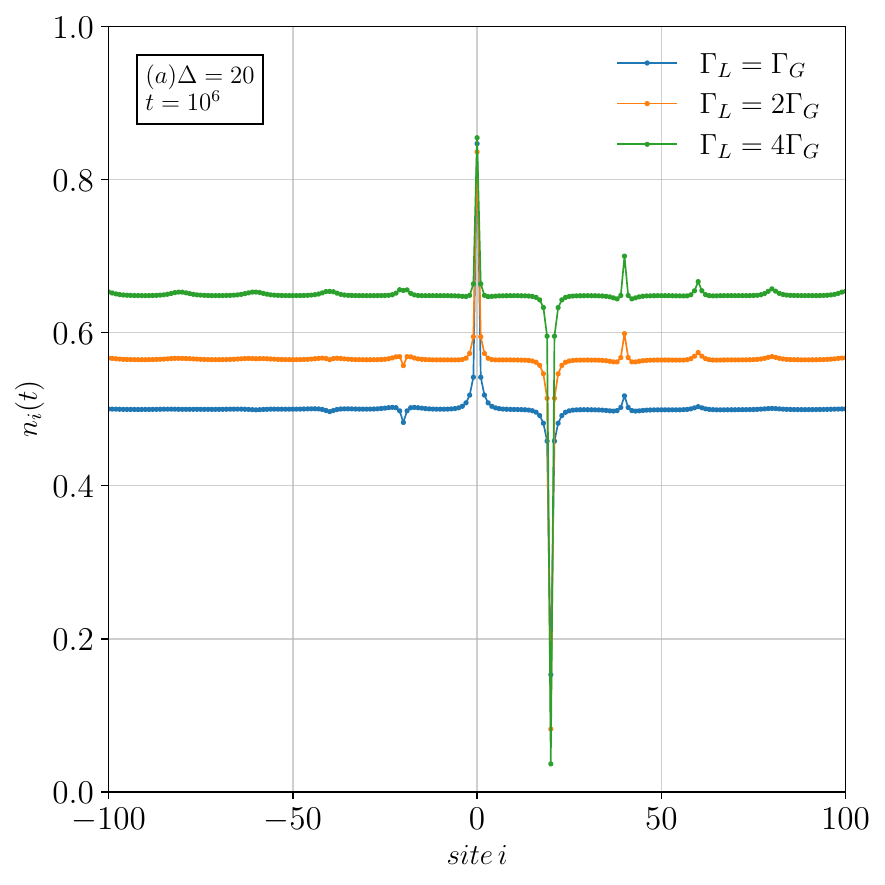}
\hfill
\includegraphics[width=0.32\textwidth]{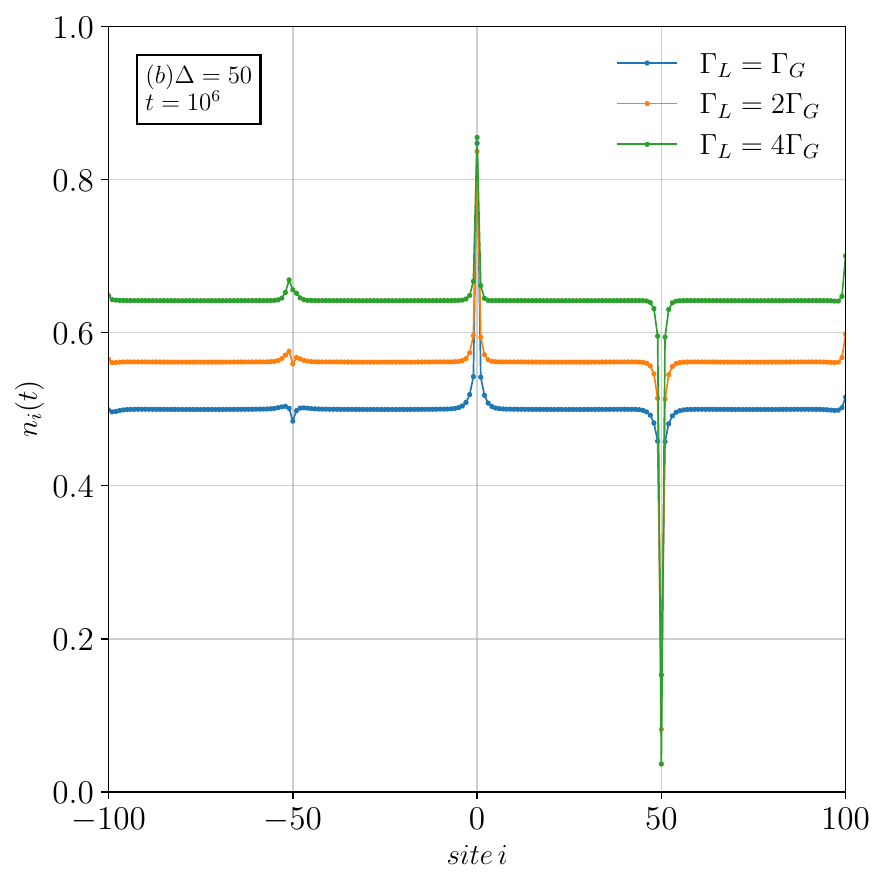}
\hfill
\includegraphics[width=0.32\textwidth]{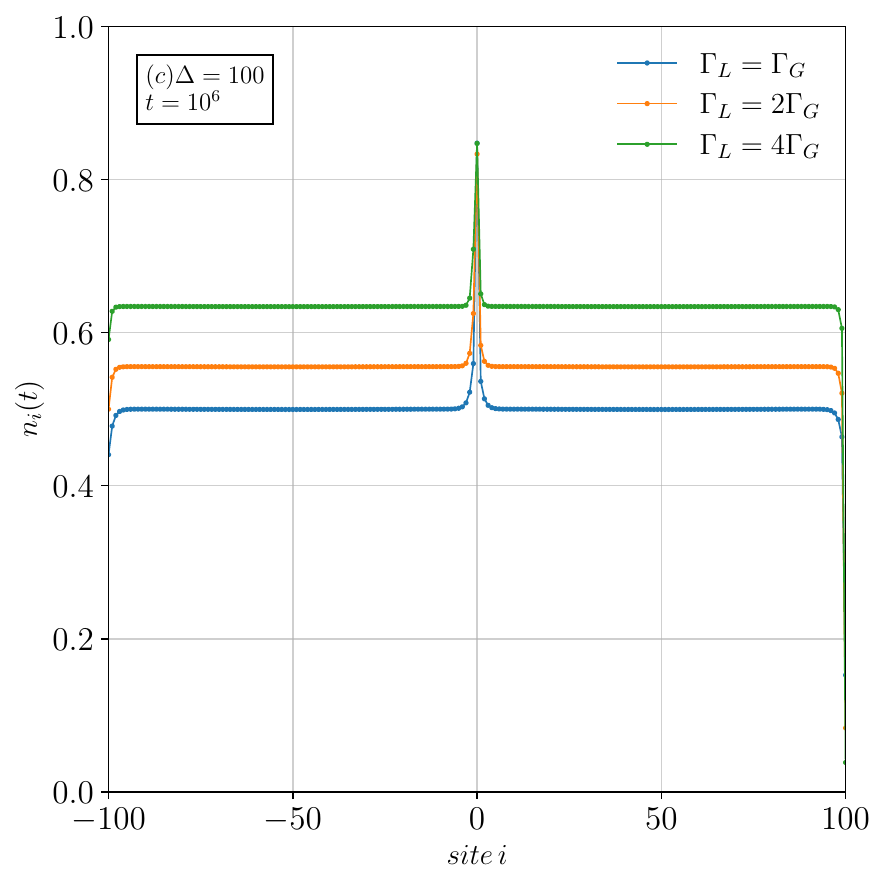}
\caption{
Local density profile $n_i(t)$ at $t = 10^6$ for $L = 201$ with source at $i = 0$ and sink at $i = \Delta$. Secondary peaks appear at distance $\Delta$ corresponding to the sink position. Parameters: $\Gamma_G = J = 1$.
}
\label{fig:diff_delta_wod2}
\end{figure*}

The results for the localized sink as well as source without dephasing are presented in Fig.~\ref{fig:diff_delta_wod2}.

%%%%%%%%%%%%%%%%%%%%%%%%%%%%%%%%%%%%%%%%%%%%%%%%%%%%%%%%%
%%%%%%%%%%%%%%%%%%%%%%%%%%%%%%%%%%%%%%%%%%%%%%%%%%%%%%%%%

\end{document}